\documentclass[preprint,showpacs,preprintnumbers,amsmath,amssymb,nofootinbib]{revtex4}

\usepackage{dcolumn}
\usepackage{bm}
\usepackage{braket}
\usepackage{amsmath}
\usepackage{txfonts}
\usepackage[pdftex]{graphicx,color}
\usepackage{float}

\begin{document}
\preprint{KUNS2609}
\preprint{KOBE-COSMO-16-03}

\title{ Chiral Primordial Gravitational Waves \\
from Dilaton Induced Delayed Chromo-natural Inflation}

\author{Ippei Obata$^{1,2}$}
\email{obata@tap.scphys.kyoto-u.ac.jp}
\author{Jiro Soda$^{2}$}
 \email{jiro@phys.sci.kobe-u.ac.jp}

\affiliation{
$^{1}$Department of Physics, Kyoto University, Kyoto, 606-8502, Japan\\
$^{2}$Department of Physics, Kobe University, Kobe, 657-8501, Japan
}

\collaboration{CLEO Collaboration}

\date{\today}

\begin{abstract}
 We study inflation driven by a dilaton and an axion,  both of which are coupled to a SU(2) gauge field.
We find that the inflation driven by the dilaton 
occurs in the early stage of inflation during which the gauge field grows due to the gauge kinetic 
function. When the energy density of magnetic fields catches up with that of electric fields,  
chromo-natural inflation takes over in the late stage of inflation, which we call delayed chromo-natural inflation. 
Thus, the delayed chromo-natural inflation driven by the axion and the gauge field is induced by
the dilaton. The interesting outcome of the model is generation of chiral primordial gravitational waves on small scales.
 Since the gauge field   is inert  in the early stage of inflation, 
 it is viable  in contrast to the conventinal chromo-natural inflation.
 We find the parameter region where chiral gravitational waves are generated in a frequency range higher than nHz, which are potentially detectable in future gravitational wave interferometers and pulsar timing arrays such as DECIGO, eLISA and SKA.
\end{abstract}

\pacs{Valid PACS appear here}

\maketitle

\tableofcontents

\section{Introduction}

 Primordial fluctuations of matter produced quantum mechanically during inflation~\cite{Guth:1980zm} 
 elegantly explains the origin of the cosmic microwave background (CMB) anisotropy 
and the large scale structure of our present universe. 
 On top of the primordial fluctuations of matter, it is well known that 
inflation quantum mechanically generates fluctuations of spacetime, so called primordial gravitational waves.
  The current  constraint on the amplitude of primordial gravitational waves in terms of the tensor-to-scalar ratio $r$ comes
 from the observation of the CMB.
 The latest joint analysis of BICEP2/Keck Array and Planck data provided an upper limit on the tensor-to-scalar ratio as ~$r ~|_{0.05} < 0.07$ ~at $95\%$ confidence level \cite{Array:2015xqh, Ade:2015lrj}.
 Remarkably, the amplitude of primordial gravitational waves is directly related to the energy scale of inflation.
The energy scale probed by the CMB observations is around the scale of grand unification theory (GUT),   $10^{16}$GeV. 
Hence, once primordial gravitational waves were detected, they would be a powerful probe of the physics in early universe
such as the GUT, supergravity, and superstring theoy. 

From the perspective of the rapid progress of cosmological observations, 
it is necessary to adapt inflation to  the percent level precision of observations.
This should be done based on fundamental theory.
As a consequence, it is expected that qualitatively new phenomena in inflation are found and those should be probed by primordial gravitational waves.
Since supergravity is the low energy limit of superstring theory and the GUT might be embedded into supergravity,
it is legitimate to investigate inflation based on supergravity. 
Indeed, there have been a lot of efforts to construct models based on supergravity~\cite{Yamaguchi:2011kg}. 
However, in the previous study, the gauge sector in supergravity has been overlooked. In fact, it was legitimate to 
truncate the gauge sector because it apparently gives rise to tiny effects on observables.
From the point of view of precision cosmology, however,
 it is worth exploring the role of the gauge sector in inflation.
 In supergravity, there exists a complex scalar field, we call its real part a dilaton and its imaginary part an axion.
 These two scalar fields are candidates for an inflaton and their couplings to the gauge sector are of our interest.
 There are two ways how an inflaton field is coupled to the gauge field.
  One is the dilaton coupling to the gauge field through a gauge kinetic function. 
 The other is an axion coupling to the gauge field. Thus, the issue to be clarified is if we can probe
  the gauge sector in inflation with the primordial gravitational waves.

 Once we introduce the dilaton field and its gauge kinetic function, inspired by supergravity, there are
  interesting phenomenologies such as the generation of primordial magnetic fields~\cite{Maleknejad:2012fw}. 
 In particular, recently, it has been found that a non-trivial gauge-kinetic function could trigger the growth of the gauge field
 which stops at the saturating point
 and the amplitude of the gauge field is sustained during inflation~\cite{Watanabe:2009ct}.
 The survived gauge field results in the statistical anisotropy in the primordial 
 fluctuations \cite{Watanabe:2010fh, Watanabe:2010bu}.
 In the Abelian gauge field cases, it accompanies anisotropic inflation~\cite{Watanabe:2009ct,Soda:2012zm}. 
In the non-Abelian gauge field cases, the isotropic 
 configuration is an attractor~\cite{Yamamoto:2012tq}. 
 However, since its convergence is quite slow, there would appear the statistical anisotropy
 depending on the initial conditions~\cite{Murata:2011wv,Maeda:2012eg}. 
Thus, we can expect the cross correlation between gravitational waves and curvature perturbations
 on top of the statistical anisotropy in auto correlation of gravitational waves~\cite{Watanabe:2010bu}.
It is also possible to generate gravitational waves from the particle production of gauge fields by the dilaton~\cite{Choi:2015wva}.

 In the presence of the axion coupling during inflation, 
  there occurs the transient tachyonic instability in one of helicity modes of the gauge field near the horizon-crossing.
 %
 Remarkably, the parity violating gauge field produces the circular polarization of gravitational waves,
 namely chiral gravitational waves~\cite{Sorbo:2011rz, Barnaby:2011vw, Cook:2011hg, Barnaby:2011qe, Anber:2012du, Barnaby:2012xt, Mukohyama:2014gba, Namba:2015gja, Dimastrogiovanni:2012ew, Adshead:2013qp, Adshead:2013nka, Obata:2014loa}.
  We can test this interesting outcome by observing the correlation between CMB temperature anisotropy and B-mode polarization \cite{Saito:2007kt, Shiraishi:2013kxa, Bartolo:2014hwa}, or by analyzing the residual signal of pulsar-timing arrays \cite{Kato:2015bye}.
 In the case of  the Abelian gauge field, however, the gauge field also produces curvature perturbations with large non-gaussianity.
  As a result, this effect easily violates the current observational bounds on the non-gaussianity of the CMB anisotropies〜\cite{ Barnaby:2010vf,Sorbo:2011rz,Barnaby:2011vw,Meerburg:2012id,Ferreira:2014zia}.
 More seriously, the strong interaction of the axion to the gauge field breaks perturbative descriptions in the in-in formalism \cite{Ferreira:2015omg}.
 These constraints severely constrain amplitudes of chiral gravitational waves and make them invisible on CMB scales.
  In the case of the non-Abelian gauge field, there is another effect of the axion coupling.
   Indeed, the interaction of the axion with the non-Abelian gauge field gives rise to an effective Hubble friction, 
and generates a slow-roll inflationary solution for a wide range of parameters~\cite{Adshead:2012kp}(see also related models~\cite{Maleknejad:2011jw}).
 In the presence of the background gauge field, the non-Abelian gauge field perturbations have  tensor perturbations and 
 one of circular polarization states experiences the tachyonic instability near the horizon crossing, which produces
  chiral gravitational waves. 
 However, this inflation model is in conflict with CMB observations because either the spectral index of the curvature perturbations
  is too red or  gravitational waves are produced too much~\cite{Dimastrogiovanni:2012ew, Adshead:2013qp, Adshead:2013nka}.
 This result stems from the fact that the strength of the interaction between the axion and the gauge field is almost constant during inflation. Hence, sizable chiral primordial gravitational waves  cannot be reconciled with the constraints on CMB scales.

 As we have explained in the above, 
 recent works have revealed roles of two types of the coupling between the inflaton and the gauge field separately.  
 From the point of view of supergravity, however,  we should incorporate  both the dilaton and the axion into a model at the same time.
 In this paper, we will focus on this possibility and find 
 a novel effect induced by the presence of both couplings, which can be probed by
 primordial gravitational waves.
 The point is that the gauge kinetic function of the dilaton 
   makes the gauge field increase and controlls   the strength of the axion-gauge field interaction.
  As initial conditions, we choose the gauge field small enough so that we can ignore an effective coupling 
  of the axion to the gauge field in the early stage of inflation. 
  Therefore, the tachyonic instability plays no role when primordial fluctuations on  CMB scales are generated. 
 However, the gauge field grows due to the dilaton coupling through the gauge kinetic function
  until the non-linear effect of gauge field appears in the background dynamics.
  Eventually, the gauge field settles in an attractor and realizes delayed chromo-natural inflation~\cite{Adshead:2012kp} 
 in the late stage of inflation, 
 Remarkably, sizable chiral gravitational waves are generated only on small scales, whose amplitude can be detectable by future space interferometers and pulsar timing arrays such as DECIGO, eLISA and SKA \cite{Seto:2001qf, AmaroSeoane:2012km, Carilli:2004nx}.

 This paper is organized as follows.
 In Sec.\ref{setup}, we present an inflation model with a dilaton and an axion, both of which are coupled with a SU(2) gauge field.
 We then derive equations of motions for the homogeneous fields and study the background dynamics.
  We show two different inflationary stages are realized in Sec.\ref{ba}.
 In Sec.\ref{pe}, first we  decompose perturbations into scalar, vector, and tensor perturbations 
 and give the gauge conditions.
 Next, we analyze tensor dynamics and show that chiral gravitational waves are produced on scales smaller than CMB scales.
 In Sec.\ref{sca}, we also discuss the dynamics of scalar perturbations.
 We estimate the amplitude of curvature perturbations on CMB scales and check the stability of scalar dynamics on small scales.
 In Sec.\ref{phe}, we discuss phenomenological predictions in this model.
 The final section is devoted to conclusion. In the Appendix A, we list equations used in the numerical calculations.

\section{Inflation model with dilaton and axion coupled to SU(2) gauge field \label{setup}}

 In this section, we present an inflationary model and derive equations of background motions.
 Specifically, we consider a dilaton field ~$\varphi$ ~and an axion field ~$\sigma$ , both of which are coupled with a SU(2) gauge field ~$A_\mu^a$ .
 The field strength of the gauge field ~$F^a_{\mu\nu}$ ~is defined by
\begin{equation}
F^a_{\mu\nu} = \partial_\mu A^a_\nu - \partial_\nu A^a_\mu + g\epsilon^{abc}A^b_\mu A^c_\nu \ ,
\end{equation}
where ~$g$ ~is its gauge coupling constant and ~$\epsilon^{abc}$ ~is the Levi-Civita symbol whose components are structure constants of SU(2) gauge field.
 The dual field strength tensor ~$\tilde{F}^{a\mu\nu}$ ~is defined by 
\begin{equation}
\tilde{F}^{a\mu\nu} = \dfrac{1}{2!}\sqrt{-g}\epsilon^{\mu\nu\rho\sigma}F^a_{\rho\sigma} \ , \qquad \epsilon^{0123} = g^{0\alpha}g^{1\beta}g^{2\gamma}g^{3\delta}\epsilon_{\alpha\beta\gamma\delta} = \dfrac{1}{-g} \ ,
\end{equation}
where ~$\epsilon_{\mu\nu\rho\sigma}$ ~is an antisymmetric tensor.
The action reads
\begin{align}
S &= S_{\text{EH}} + S_{\text{dilaton}} + S_{\text{axion}} + S_{\text{gauge}} + S_{\text{CS}} \notag \\
   &=\int dx^4\sqrt{-g}\left[\dfrac{1}{2}R-\dfrac{1}{2}(\partial_\mu\varphi)^2 -V(\varphi) -\dfrac{1}{2}(\partial_\mu\sigma)^2 -W(\sigma) -\dfrac{1}{4}I(\varphi)^2F^{a\mu\nu}F^a_{\mu\nu}-\dfrac{1}{4}\lambda\dfrac{\sigma}{f}\tilde{F}^{a\mu\nu}F^a_{\mu\nu} \right] \label{eq:action} \ ,
\end{align}
where we used units ~$\hbar=c=1$ ~and ~$M_{\text{pl}}=(8\pi G)^{-1/2}=1$ .
 Here, ~$g$ ~is a determinant of a metric ~$g_{\mu\nu}$ ~(note that it is not related with the gauge coupling constant), ~$R$ ~is a Ricci scalar,
 and ~$f$ ~is a decay constant of axion.
 Moreover, we introduced a coupling constant of axion to the gauge field ~$\lambda$ .
We have also introduced the potential functions $V(\varphi)$ and $W(\sigma)$.

As to the metric, we adapt the ADM parametrization
\begin{equation}
ds^2=-N^2dt^2+q_{ij}(dx^i+N^idt)(dx^j+N^jdt) \ ,
\end{equation}
where ~$N$ ~is a lapse function and ~$N^i$ ~is a shift function, which are Lagrange multipliers of the system.
 An induced metric $q_{ij}$ on the three dimensional spatial hypersurface  is used to raise or lower the index as 
\begin{equation}
q^{ik}q_{kj}=\delta^i_j \ , \qquad N^i=q^{ij}N_j \ .
\end{equation}
Thus, the metric can be expressed by
\begin{equation}
g_{\mu\nu} = \left( \begin{array}{cc}
-N^2 + N_i N^i & N_j \\
 N_i  & q_{ij}
\end{array} \right) \ , \qquad g^{\mu\nu} = \left( \begin{array}{cc}
-N^{-2} & N^{-2}N^j \\
 N^{-2}N^i  & q^{ij} - N^{-2}N^iN^j
\end{array} \right) \ .
\end{equation}
 Using these variables, we can rewrite the action \eqref{eq:action} as
\begin{align}
S_{\text{EH}} &=\int dx^4N\sqrt{q}\left[\dfrac{1}{2}\left({^{(3)}}R+K_{ij}K^{ij}-K^2\right)\right] \ , \\
S_{\text{dilaton}} &= \int dx^4N\sqrt{q}\left[\dfrac{1}{N^2}\left(\dfrac{1}{2}\dot{\varphi}^2-N^i\dot{\varphi}\varphi_{,i}+\dfrac{1}{2}(N^i\varphi_{,i})^2\right)-\left(\dfrac{1}{2}q^{ij}\varphi_{,i}\varphi_{,j}+V(\varphi)\right)\right] \ , \\
S_{\text{axion}} &= \int dx^4N\sqrt{q}\left[\dfrac{1}{N^2}\left(\dfrac{1}{2}\dot{\sigma}^2-N^i\dot{\sigma}\sigma_{,i}+\dfrac{1}{2}(N^i\sigma_{,i})^2\right)-\left(\dfrac{1}{2}q^{ij}\sigma_{,i}\sigma_{,j}+W(\sigma)\right)\right] \ , \\
S_{\text{gauge}} &= \int dx^4N\sqrt{q}\left[ \dfrac{1}{2N^2}I(\varphi)^2q^{ik}(F^a_{0i}+F^a_{ij}N^j)(F^a_{0k}+F^a_{kl}N^l)-\dfrac{1}{4}I(\varphi)^2q^{ik}q^{jl}F^a_{ij}F^a_{kl}\right] \ , \\
S_{\text{CS}} &= \int dx^4\left[-\dfrac{1}{2}\lambda\dfrac{\sigma}{f}\epsilon^{ijk}F^a_{0i}F^a_{jk}\right] \ .
\end{align}
 Note that Chern-Simons interaction does not include metric variables.
 Here, ~$q$ ~is a determinant of the metric ~$q_{ij}$. We defined the extrinsic curvature  of a hypersurface
\begin{eqnarray}
K_{ij} \equiv \dfrac{1}{2N}(\dot{q}_{ij} - 2N_{(i|j)}) \ , 
\end{eqnarray}
and the Ricci scalar of the hypersurface
\begin{eqnarray}
{^{(3)}}R = (q_{ij,kl}+q_{mn}{^{(3)}}\Gamma^m_{ij}{^{(3)}}\Gamma^n_{kl})(q^{ik}q^{jl}-q^{ij}q^{kl}) \ ,
\end{eqnarray}
where
\begin{equation}
{^{(3)}}\Gamma^i_{jk} = \dfrac{1}{2}q^{il}(q_{lj,k}+q_{lk,j}-q_{jk,l}) \ .
\end{equation}

 Let us consider the homogeneous background dynamics in this set-up.
For the metric, we use a spatially flat metric
\begin{equation}
N=N(t) \ , \quad N_i=0 \ , \quad q_{ij}=a(t)^2\delta_{ij} \ .
\end{equation}
 After taking the variation of the action, we set ~$N(t)=1$. Thus, the time function ~$t$ becomes the cosmic time.
 For the dilaton and the axion we take homogeneous configurations ~$\varphi = \varphi(t) \ ,  \sigma = \sigma(t)$.
 As a gauge condition, we choose the temporal gauge
\begin{equation}
A^a_0 = 0 \ .
\end{equation}
We also take an ansatz
\begin{equation}
A^a_i=A(t)\delta^a_i=a(t)Q(t)\delta^a_i \label{eq: defgauge} \ ,
\end{equation}
which is invariant under the diagonal transformation of the spatial rotation SO(3) and the SU(2) gauge symmetry.
 Note that ~$Q(t)$ ~is a scalar under the diagonal transformation and we use this variable later.
 Thus, their field strength ~$F^a_{\mu\nu}$ ~can be deduced as
\begin{equation}
F^a_{0i}=\dfrac{dA}{dt}\delta^a_i \equiv aE(t)\delta^a_i \ , \qquad F^a_{ij}=g\epsilon^{abc}A^b_iA^c_j=gA^2\epsilon^a_{ij} \equiv a^2B(t)\epsilon^a_{ij} \ ,
\end{equation}
where we defined electric and magnetic components ~$E(t) \ , B(t)$ .
 Substituting these configurations into the action, we obtain the following background action
\begin{equation}
S=\int d^4x\dfrac{a^3}{N}\left[ -3\dfrac{\dot{a}^2}{a^2}+\dfrac{1}{2}\dot{\varphi}^2 - N^2V +\dfrac{1}{2}\dot{\sigma}^2 - N^2W +\dfrac{3}{2}I^2\left(E^2 - N^2B^2 \right) - 3N\dfrac{\lambda}{f}\sigma E B \right] \ ,
\end{equation}
where a dot denotes a derivative with respect to the cosmic time $t$ .
 Taking the variation with respect to ~$N$ and setting $N=1$ after the variation,  we obtain the Hamiltonian constraint
\begin{equation}
3H^2=\dfrac{1}{2}\dot{\varphi}^2+V+\dfrac{1}{2}\dot{\sigma}^2+W+\rho_E+\rho_B \label{eq: Friedmann} \ ,
\end{equation}
where ~$H \equiv \dot{a}/a$ ~is the Hubble parameter.
 Here, we defined the following energy densities of electric and magnetic fields
\begin{equation}
\rho_E \equiv \dfrac{3}{2}I^2E^2 = \dfrac{3}{2}I^2\dfrac{\dot{A}^2}{a^2} \ , \qquad \rho_B \equiv \dfrac{3}{2}I^2B^2 = \dfrac{3}{2}I^2\dfrac{g^2A^4}{a^4} \ .
\end{equation}
 The equations for the dilaton, the axion and the gauge field read
\begin{gather}
\ddot{\varphi} + 3H\dot{\varphi} + V_{\varphi} = 2\dfrac{I_\varphi}{I}\left( \rho_E - \rho_B \right) \ , \label{eq: dilamotion} \\
\ddot{\sigma} + 3H\dot{\sigma} + W_{\sigma} = - 3\dfrac{\lambda}{f}E B \ , \label{eq: aximotion} \\
\ddot{A}+\left(H+2\dfrac{\dot{I}}{I}\right)\dot{A}+2g^2\dfrac{A^3}{a^2} = \dfrac{\lambda}{f}\dot{\sigma}g\dfrac{A^2}{aI^2} \label{eq: gaugemotion} \ ,
\end{gather}
where
\begin{equation}
V_\varphi \equiv \dfrac{dV}{d\varphi} \ , \quad W_\sigma \equiv \dfrac{dW}{d\sigma} \ , \quad I_\varphi \equiv \dfrac{dI}{d\varphi} \ .
\end{equation}
 The equation for the scale factor $a(t)$ is given by
\begin{equation}
\dot{H} = - \left(\dfrac{1}{2}\dot{\varphi}^2 + \dfrac{1}{2}\dot{\sigma}^2 + \dfrac{2}{3}(\rho_E+\rho_B) \right) \ .
\end{equation}
Now that we have obtained the basic equations, we can discuss the inflationary dynamics.

\subsection{Inflationary dynamics \label{ba}}

 In this section, we analyze the background dynamics which eventually 
produces chiral gravitational waves on small scales.

 First, we discuss initial conditions.
 We assume that the dilaton is energetically dominant and plays a role of an inflaton in the early stage of inflation.
 The dynamics during inflation can be described by  slow-roll equations
\begin{equation}
3H^2 \simeq V \ , \qquad 3H\dot{\varphi} + V_\varphi \simeq 0 \label{eq: slow-roll1} \ .
\end{equation}
 In addition to those, we assume that the axion also satisfies a slow-roll equation
\begin{equation}
3H\dot{\sigma} + W_{\sigma} \simeq 0 \ .
\end{equation}
Now, we introduce the slow-roll parameters in this system
\begin{align}
\epsilon_H &\equiv -\dfrac{\dot{H}}{H^2} = \epsilon_\varphi + \epsilon_\sigma + \epsilon_E + \epsilon_B \ , \\
\eta_H &\equiv \dfrac{\dot{\epsilon}_H}{H\epsilon_H} = 2\left( \epsilon_H + \dfrac{\epsilon_\varphi}{\epsilon_H}\dfrac{\ddot{\varphi}}{H\dot{\varphi}} + \dfrac{\epsilon_\sigma}{\epsilon_H}\dfrac{\ddot{\sigma}}{H\dot{\sigma}} + \dfrac{\epsilon_E}{\epsilon_H}\left( \dfrac{\dot{I}}{HI} + \dfrac{\dot{E}}{HE} \right) +\dfrac{\epsilon_B}{\epsilon_H}\left( \dfrac{\dot{I}}{HI} + \dfrac{\dot{B}}{HB} \right) \right) \ ,
\end{align}
where we defined  slow-roll parameters
\begin{align}
\epsilon_\varphi \equiv \dfrac{1}{2}\dfrac{\dot{\varphi}^2}{H^2} \ , \quad \epsilon_\sigma \equiv \dfrac{1}{2}\dfrac{\dot{\sigma}^2}{H^2} \ , \quad \epsilon_E \equiv  \dfrac{2}{3}\dfrac{\rho_E}{H^2} \ , \quad \epsilon_B \equiv  \dfrac{2}{3}\dfrac{\rho_B}{H^2} \ .
\end{align}
 Here, we assume the relation ~$\epsilon_\sigma \ll \epsilon_H$ which can be realized easily.
 We can express the scale factor as a function of $\varphi$ :
\begin{align}
a &\simeq a_0\exp[-\int_0\dfrac{V}{V_\varphi}d\varphi] = a(\varphi) \  ,
\end{align}
where the index ``0" represents an initial value.
 Inspired by this result,  we put the gauge kinetic function as
\begin{equation}
I(\varphi) = I_0\exp[-n \int_0\dfrac{V}{V_\varphi}d\varphi] \label{eq: I} \ ,
\end{equation}
where ~$I_0$ ~is a constant.
 Thus, we have an approximate relation ~$I(\varphi) \propto a^n$ during slow-roll inflation.
Note that ~$n$ ~is a parameter and it controls the strength of the gauge coupling.

In this paper, we consider a gauge field with a weak coupling constant.
 Namely, ~$I(\varphi)$ ~is a decreasing function ($n<0$) ~and we choose its initial value ~$I_0$ ~very large in order to get ~$g^{-1}I \sim \mathcal{O}(1)$ ~at the end of inflation.
 This assumption is necessary to avoid the strong coupling problem.
 Hence the electric and magnetic components must be very small in order to satisfy slow-roll equations \eqref{eq: slow-roll1}.
 The naturalness of this requirement depends on the effective potential of the gauge field, as we discuss later.
 Due to this assumption, we can neglect their non-linear effects at first.
 Then, Eq.\eqref{eq: gaugemotion} can be reduced to
\begin{equation}
\ddot{A}+\left(H+2\dfrac{\dot{I}}{I}\right)\dot{A} \simeq 0 \ ,
\end{equation}
which can be integrated as
\begin{equation}
     aI^2\dot{A} = \text{const.} \equiv C \label{eq: slow-gauge} \ .
\end{equation}
 Thus, the energy density of electric field evolves as
\begin{equation}
\rho_E = \dfrac{3C^2}{2a^4I^2} \propto a^{-2(n+2)} \label{eq: rhoE} \ .
\end{equation}
 Since we are interested in the solution where the energy density of the gauge field does not  decay, 
 we consider ~$n\leq-2$ ~region.
 On the other hand, the growing mode of magnetic energy density is proportional to ~$a^{-2(3n+4)}$ ~so that ~$\rho_B$ ~grows more rapidly than ~$\rho_E$ .
 Therefore, we can separate the inflationary history of this model into two stages.
 In the early stage of inflation, ~$\rho_B$ ~is negligibly small compared to $\rho_E$ so its effect on the dynamics 
can be ignored in the background motion.
 We assume that  primordial fluctuations on CMB scales are generated during this stage.
 In the late stage of inflation, however, ~$\rho_B$ ~grows sufficiently first and catches up with ~$\rho_E$ .
 As we explicitly show later, during this stage, the interaction between the gauge field and the axion become important and
 an inflationary dynamics goes into an attractor solution, which is quite similar to the dynamics of chromo-natural inflation \cite{Adshead:2012kp} .
 From now on, we explain these two inflationary stages separately.

\subsubsection{The early stage of inflation \label{inea}}

 In an initial period, the energy density of the gauge field is so small that only a dilaton contributes an inflationary dynamics.
 However, when ~$\rho_E$ ~grows sufficiently, the slow-roll condition for ~$\varphi$ ~will be no more valid and modified as
\begin{equation}
3H\dot{\varphi} + V_\varphi \simeq 2\dfrac{I_\varphi}{I}\rho_E \label{eq: slow2} \ .
\end{equation}
 Substituting \eqref{eq: rhoE} into \eqref{eq: slow2} , we can  solve this equation~\cite{Watanabe:2009ct} and find the gauge kinetic function as
\begin{equation}
a^4I^2 \simeq -\dfrac{n^2}{n+2}\dfrac{3C^2}{2\epsilon_V V} ( 1 + D_1 a^{2(n+2)} ) 
         \label{eq: Ia}   \ ,
\end{equation}
 where ~ $ D_1 $ is a constant of integration
   and $\epsilon_V \equiv \frac{1}{2}(\frac{V_\varphi}{V})^2$ ~is the slow-roll parameter in terms of the potential.
Here, we note the relation
\begin{eqnarray}
\epsilon_E \simeq -\dfrac{2(n+2)}{n^2}\epsilon_V  \ .
\end{eqnarray}
 We neglected the time dependence of $\epsilon_V V$ in the above.
 We can see that the second term in \eqref{eq: Ia} becomes soon negligible when $n < -2$. 
Thus, ~$\rho_E$ ~settles in a nearly constant value during this period
\begin{equation}
\rho_E \simeq -\dfrac{n+2}{n^2}\epsilon_V V   \ .
\end{equation}
That implies the effective $n$ goes to $-2$~\cite{Kanno:2009ei}.
 At this time the relation between ~$\epsilon_V$ ~and ~$\epsilon_\varphi$ ~reads
\begin{equation}
\sqrt{\epsilon_\varphi} \simeq \sqrt{\epsilon_V} + \dfrac{n\epsilon_E}{\sqrt{2\epsilon_V}} \label{slowV} \ .
\end{equation}
 Remarkably, an electric component of gauge field finally supports  the slow-roll inflation in this period.
 Then Hubble slow-roll parameters read
\begin{align}
\epsilon_H &\simeq \epsilon_\varphi + \epsilon_E \label{slowH} \ , \\
\eta_H &\simeq 2\left( \epsilon_H + \dfrac{\epsilon_\varphi}{\epsilon_H}\dfrac{\ddot{\varphi}}{H\dot{\varphi}} + \dfrac{\epsilon_E}{\epsilon_H}\left( \dfrac{\dot{I}}{HI} + \dfrac{\dot{E}}{HE} \right) \right) \ .
\end{align}

\subsubsection{The late stage of inflation}

 Up to here, we have neglected the  non-linear effect of the gauge field.
 However,  the energy density of magnetic field ~$\rho_B$ ~grows and eventually becomes
 greater than ~$\rho_E$ for $n\leq-2$.
 As the non-linear effect becomes important,
  we cannot use Eq.\eqref{eq: slow-gauge}. Instead, we have to consider the full equation of motion for the gauge field \eqref{eq: gaugemotion}.
 Here, we rewrite Eq.\eqref{eq: gaugemotion} into the equation for ~$Q$ ~defined by \eqref{eq: defgauge}
\begin{equation}
\ddot{Q} + \left(1+\dfrac{2}{3}\dfrac{\dot{I}}{HI}\right)3H\dot{Q} = - \left( \left(2 - \epsilon_H + 2\dfrac{\dot{I}}{HI}\right)H^2 Q + 2g^2Q^3 - \dfrac{\lambda}{f}\dot{\sigma}\dfrac{gQ^2}{I^2} \right) \label{eq: phimotion} \ ,
\end{equation}
which looks like an equation for a scalar field.
 Hence we can interpret the right hand side of Eq.\eqref{eq: phimotion} as an effective potential force of the gauge field.
It is easy to read off the effective potential from Eq.\eqref{eq: phimotion} as
\begin{equation}
U_{\text{eff}}(Q) = \left(1+\dfrac{\dot{I}}{HI}\right)H^2Q^2 -\dfrac{1}{3}\dfrac{\lambda}{f}\dot{\sigma}\dfrac{gQ^3}{I^2} + \dfrac{1}{2}g^2Q^4 \label{eq: Ueff} \ .
\end{equation}
 Note that we neglected the slow-roll corrections.

 We can understand a background dynamics of the gauge field from this effective potential.
 Let us focus on the coefficient of ~$Q^2$ ~term.
 Before inflation occurs, we assume that its coefficient was positive and ~$Q$ ~is near an origin of its potential.
 However, we assume that at a certain time ~$\frac{\dot{I}}{HI} \simeq n \leq -2$ ~is realized and ~$Q$ ~starts to roll down on the potential.
 Even if ~$\rho_E$ ~modifies slow-roll dynamics, ~$\frac{\dot{I}}{HI} = -2 + \mathcal{O}(\epsilon_H)$ ~holds due to \eqref{eq: Ia} .
 Therefore, in an early period ~$Q$ ~grows up because of the negativity of ~$Q^2$ ~term.
 However, ~$Q$ ~stops its growth when non-linear magnetic ~$Q^4$ ~term becomes important. 
 Moreover, due to axion-gauge interaction ~$Q^3$ ~term, this effective potential gets two different minimum values, so we expect that there is a trajectory where the gauge field finally settles in a deeper bottom of this potential.
 At this time, $\sigma$ and $Q$ obey slow-roll equations
\begin{gather}
3H\dot{\sigma} + W_{\sigma} = - 3\dfrac{\lambda}{f}(\dot{Q} + HQ)gQ^2 \ , \\
\left(1+\dfrac{2}{3}\dfrac{\dot{I}}{HI}\right)3H\dot{Q} + 2\left(1 + \dfrac{\dot{I}}{HI}\right)H^2 Q + 2g^2Q^3 = \dfrac{\lambda}{f}\dot{\sigma}\dfrac{gQ^2}{I^2} \label{eq: phislowmotion} \ .
\end{gather}
 Diagonalizing this system, we have
\begin{gather}
\left( 1 + \dfrac{2}{3}\dfrac{\dot{I}}{HI} + \dfrac{1}{3}\Lambda^2m_Q^2 \right)3H\dot{\sigma} + \left( 1 + \dfrac{2}{3}\dfrac{\dot{I}}{HI} \right)\left( W_\sigma + 3\dfrac{\lambda}{f}HgQ^3 \right) - 2\left(1 + \dfrac{\dot{I}}{HI} + m_Q^2 \right)\dfrac{\lambda}{f}HgQ^3 = 0 \ , \\
\left( 1 + \dfrac{2}{3}\dfrac{\dot{I}}{HI} + \dfrac{1}{3}\Lambda^2m_Q^2 \right)3H\dot{Q} + \left( 1 + \dfrac{\dot{I}}{HI} + \left(\dfrac{1}{2}\Lambda^2+1\right)m_Q^2 \right)2H^2Q + \dfrac{1}{3I}\Lambda m_Q W_\sigma = 0 \ ,
\end{gather}
where we defined two model parameters
\begin{equation}
\Lambda \equiv \dfrac{\lambda}{I f}Q \ , \qquad m_Q \equiv \dfrac{g Q}{H} \ .
\end{equation}
 If ~$\Lambda^2 \gg 1$ ~and ~$m_Q^2 \gg \Lambda^{-2}$ ~hold, we get the following results
\begin{align}
Q &\simeq Q_{\text{min}} = -\left( \dfrac{f W_\sigma}{3\lambda g H} \right)^{1/3} \ , \\
\dfrac{1}{2}\dfrac{\lambda}{I^2}\dfrac{\dot{\sigma}}{f H} &\simeq m_Q + \dfrac{1}{m_Q}\left( 1 + \dfrac{\dot{I}}{HI} \right) \label{eq: barxi} \ .
\end{align}
 Remarkably, the background gauge field settles in a minimum value of its effective potential.
 This is essentially an attractor solution in chromo-natural inflation \cite{Adshead:2012kp}.
 In our model, however, the condition ~$\Lambda \gg 1$ ~and the  slow-roll equation \eqref{eq: barxi} can be controlled by the gauge kinetic function.
 So we can realize this slow-roll trajectory for a broad parameter region compared to the original chromo-natural inflation.
 Interestingly, this inflationary dynamics generates parity-violating chiral gravitational waves.
 We discuss the detail of this mechanism later.

 Finally, let us check the evolution of the gauge kinetic function during this period.
 In this period, the energy densities of the electric and the magnetic fields are given by
\begin{equation}
\rho_E \simeq \dfrac{3}{2}I^2H^2 Q_{\text{min}}^2 \ , \quad \rho_B \simeq \dfrac{3}{2}I^2g^2 Q_{\text{min}}^4 = m_Q^2~\rho_E \label{eq: ene} \ .
\end{equation}
 Hence if ~$m_Q \gtrsim 1$ ~holds, the slow-roll equation for $\varphi$ reads
\begin{equation}
3H\dot{\varphi} + V_{\varphi} \simeq 2\dfrac{I_\varphi}{I}\left( \rho_E - \rho_B \right) \ .
\end{equation}
 Substituting \eqref{eq: ene} into the above equation, we obtaine the time evolution of ~$I(\varphi)$ as
\begin{equation}
I \simeq \left[ \dfrac{3H^2 Q_{\text{min}}^2}{4\epsilon_V V}(m_Q^2 - 1) + D_2 ~a^{-2nV/3H^2} \right]^{-1/2} \ , \label{eq: Ib}
\end{equation}
where $D_2$ is a constant of integration.
 Note that we neglected the terms which are suppressed in the slow-roll approximation.
 The first term in the parenthesis is almost constant. 
 Moreover, when the energy density of the axion becomes dominant in the late stage of inflation $3H^2 \simeq W \gg V$ , 
 The second term in the parenthesis is also almost constant.
 Hence, $I$ decreases very slowly.

\subsection{Numerical analysis}

As a numerical example, let us show this background dynamics by using the following potentials
\begin{align}
V(\varphi) &= \Lambda_\varphi^4 \exp \left[ r\varphi \right] \ , \\
W(\sigma) &= \Lambda_\sigma^4 \left[ 1 - \cos \left( \dfrac{\sigma}{f} \right) \right] \ ,
\end{align}
where ~$\Lambda_\varphi$ ~and ~$\Lambda_\sigma$ characterize energy densities of the dilaton and the axion.
Here, $f$ is the decay constant and  $r$ is a parameter.
It is well known that single-field power-law inflation is in conflict with CMB observations.
 However, in this model the background gauge field modifies the slow-roll dynamics and consequently
 the tensor-to-scalar ratio is suppressed. Hence, there is a room the model becomes viable  as we will discuss later.

\begin{figure}[htbp]
\begin{tabular}{cc}
\begin{minipage}{0.5\hsize}
\begin{center}
\includegraphics[width=\hsize]{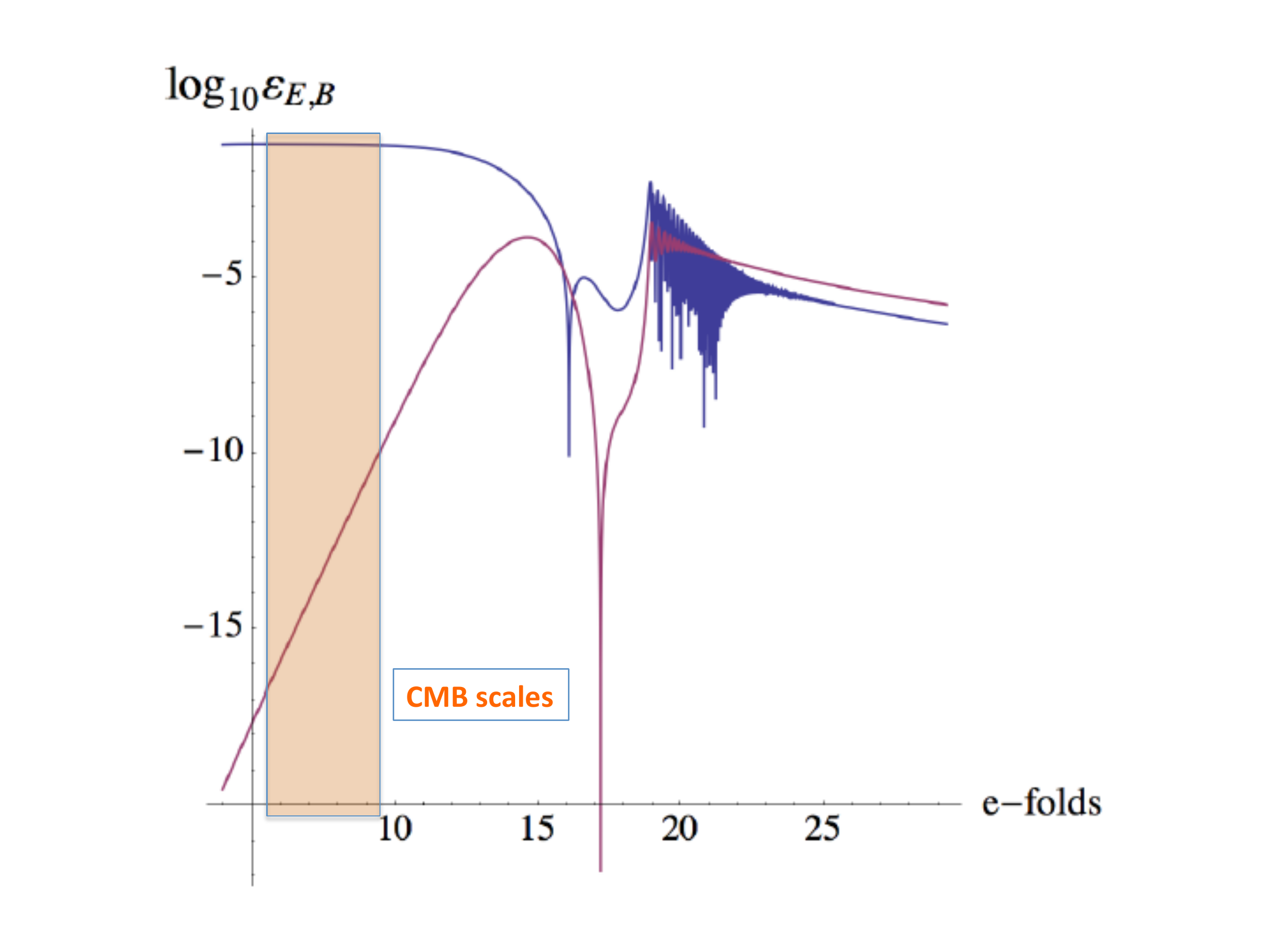}
\end{center}
\end{minipage}
\begin{minipage}{0.5\hsize}
\begin{center}
\includegraphics[width=\hsize]{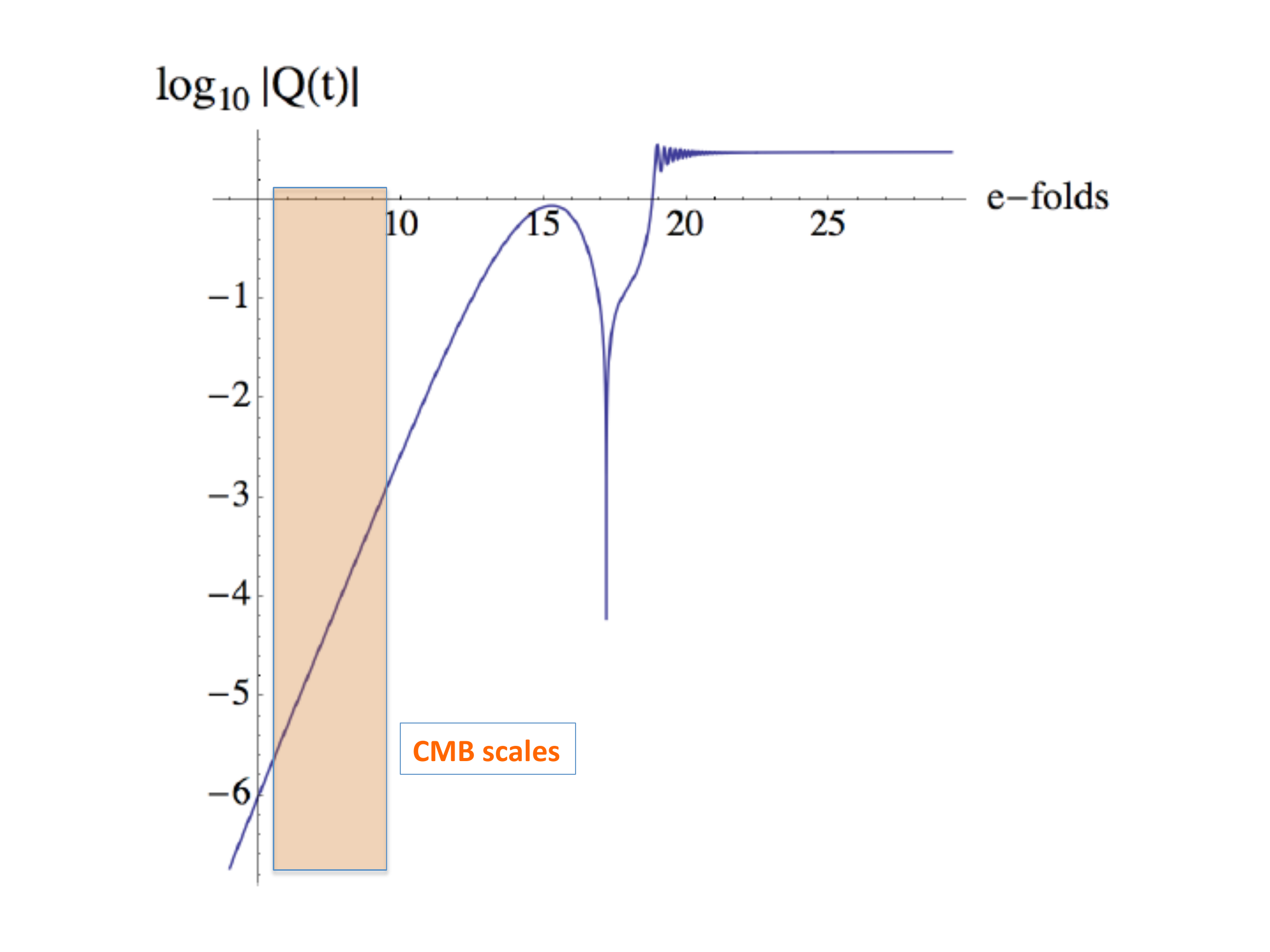}
\end{center}
\end{minipage}
\end{tabular}
\caption{(left figure) : We plotted time evolutions  ~$\epsilon_E$ ~(blue line) and ~$\epsilon_B$ ~(red line).
 Around at 15 e-folds, ~$\rho_B$ ~catches up with ~$\rho_E$ and the transition of the dynamics of the gauge field happens 
due to the non-linear effect.
 After 20 e-folds, the gauge field settles in an attractor and both $\epsilon_E$  and ~$\epsilon_B$  decrease as ~$I^2$.
 Note that the period when primordial CMB fluctuations are generated is about from 5 to 10 e-folds in this plot.
(right figure) : The time evolution of the amplitude of the gauge field ~$Q(t)$ . At first it continues to grow until 15 e-folds.
 After the transition, it settles in a constant value.
 In both figures, we used initial conditions ~$(\varphi_0 \ , \sigma_0/f) = (4 \ , \pi/3)$ and the parameters $( \ \Lambda_\varphi, \ \Lambda_\sigma, \ r, \ n, \ g, \ \lambda, \ f \ ) = ( \ 10^{-2}, \ 2\times10^{-3}, 1, \ -2.01, \ 10^{-6}, \ 10^{-1}, \ 10 \ )$.}
\label{fig: gauge}
\end{figure}

 In FIG.\ref{fig: gauge}, we plotted the time evolution of the energy density of the gauge field
in terms of ~$\epsilon_E$ ~and ~$\epsilon_B$, and the time evolution of the gauge field ~$Q(t)$.
 As you can see, ~$\rho_B$ ~catches up with ~$\rho_E$ after $15$ e-folds and the transition of the dynamics of the gauge field happens due to the non-linear effect.
 We can also see that ~$Q(t)$ ~stops its growth and oscillates 
when the magnetic energy density grows sufficiently, and finally settles in a constant value.
 Note that we used a super-Planckian decay constant ~$f > M_{\text{pl}}$ in the plot by assuming it 
can be effectively produced by the combination of sub-Planckian decay constants derived from  aligned multiple axions \cite{Kim:2004rp}.

In the next section, we show that, in the late stage of inflation, 
 one of circular polarization states of gauge field fluctuations are enhanced due to the tachyonic instability and
 the rapid growth of the gauge field produces sizable circularly polarized  gravitational waves.

\section{Chiral gravitational waves \label{pe}}

 In this section, we consider a perturbed universe. To analyze the dynamics, we need to define perturbed quantities.
 First, for the metric we use the following variables
\begin{equation}
N=1+2\phi \ , \quad N_i=\partial_i ~\beta + \beta_i \ , \quad q_{ij}=a(t)^2(\delta_{ij}+\gamma_{ij}) \label{eq: metper}  \ ,
\end{equation}
where the spatial metric is further decomposed as
\begin{equation}
\gamma_{ij} = -2\psi\delta_{ij} + 2E_{,ij} + 2W_{(i,j)} + h_{ij} \ .
\end{equation}
 Note that ~$\beta_i$ ~and ~$W_i$ ~are transverse vectors and ~$h_{ij}$ ~is a transverse traceless tensor.
 We choose the flat-slicing gauge
\begin{equation}
\psi = E = W_i = 0 \ .
\end{equation}
 Next, we decompose the non-Abelian gauge field as
\begin{align}
A^a_0 &= a(t)\left[\partial_a Y+Y_a\right] \ , \\
A^a_i &= a(t)\left[(Q(t)+\delta Q) \delta_{ai} + \epsilon_{iac}(\partial_c U+U_c) + \partial_i(\partial_a M+M_a)+T_{ai}\right] \ ,
\end{align}
where ~$Y_a \ , U_a$ ~and ~$M_a$ ~are transverse vectors and ~$T_{ai}$ ~is a transverse traceless tensor.
 Here, we did not discriminate between  the index $a$ and the spatial index $i$.
 This is allowed because the diagonal transformation of ~SU(2) and the rotation ~SO(3) remains the symmetry in the background. Remarkably, not only scalar and vector but also tensor perturbations exist.
 As is usual, scalar, vector, and tensor perturbations  are decouple at the linear level.
 We will see that ~$Y$ ~and ~$Y_a$ ~are non-dynamical variables.

 Due to SU(2) gauge symmetry,  there is the following gauge transformation
\begin{equation}
 A^a_\mu \longrightarrow A^a_\mu + g^{-1}\partial_\mu \alpha^a + \epsilon^{abc}A^b_\mu \alpha^c \ ,
\end{equation}
where $\alpha^{a}$ are  gauge parameters.
We can eliminate three degrees of freedom  using this gauge freedom.
 So, we fix its gauge by setting
\begin{equation}
M=M_a=0 \ .
\end{equation}
 Finally, we decompose (pseudo) scalar fields into their background and perturbation variables
\begin{equation}
\varphi=\bar{\varphi}(t) + \delta\varphi \ , \qquad \sigma=\bar{\sigma}(t) + \delta\sigma \ .
\end{equation}

\subsection{Tensor perturbation dynamics \label{ten}}

 Let us analyze the dynamics of tensor perturbations in this model.
 Using a new variable ~$\psi_{ij}\equiv a(\tau)h_{ij}$, the quadratic action ~$S_{\text{EH}}$ ~for tensor perturbations is given by
\begin{equation}
\delta S_{\text{EH}}=\int d\bm{x}d\tau \dfrac{1}{2}\left[\dfrac{1}{4}\psi'^{ij}\psi'_{ij}-\dfrac{1}{4}\psi^{ij,k}\psi_{ij,k}-\left(\dfrac{3}{4}\dfrac{a''}{a}-\dfrac{1}{2}\left(\dfrac{a'}{a}\right)^2\right)\psi^{ij}\psi_{ij}\right] \ ,
\end{equation}
where a prime represents a derivative with respect to a conformal time ~$\tau$ .
 We also have  contributions to the quadratic action for tensor perturbations from the (pseudo) scalar actions as
\begin{equation}
\delta S_{\text{dilaton}}+\delta S_{\text{axion}}=\int d\bm{x}d\tau \left(-\dfrac{a^2}{4}\psi^{ij}\psi_{ij}\right)\left[ \dfrac{1}{2a^2}(\bar{\varphi}'^2+\bar{\sigma}'^2) - V(\bar{\varphi})-W(\bar{\sigma}) \right] \ .
\end{equation}
 Moreover, using a new variable ~$t_{ai} \equiv a(\tau)I(\bar{\varphi})T_{ai}$ , 
 we obtain the quadratic actions for the tensor perturbations
 from the gauge  sector 
\begin{eqnarray}
\delta S_{\text{gauge}} &=& \int d\bm{x}d\tau\left(-\dfrac{1}{4}\right)\left[ -I^2\dfrac{(a Q)'^2}{2a^2}\psi^{ij}\psi_{ij} + \dfrac{3}{2}a^2I^2g^2Q^4\psi^{ij}\psi_{ij} - 2\left( t'^a_i t'^a_i + \dfrac{I''}{I}t^a_i t^a_i \right)  + 2t^a_{i,j}t^a_{i,j} \right. \notag \\
&&\left. - 4a g Q\epsilon^{abi}t^b_jt^a_{j,i} + \dfrac{4I}{a}(aQ)' \psi^{ij}\left( t'_{ij}-\dfrac{I'}{I}t_{ij} \right) - 4a I g Q^2\psi^{jm}\epsilon^a_{ij}t^a_{m,i}-4a^2I g^2Q^3\psi^{ij}t_{ij} \right] 
\end{eqnarray}
and  that from the Chern-Simons sector 
\begin{eqnarray}
\delta S_{\text{CS}} = \int d\bm{x}d\tau\dfrac{1}{2I^2}\dfrac{\lambda}{f}\bar{\sigma}' \left(\epsilon^{ijk}t^a_i t^a_{k,j} - a g Q t^{ij}t_{ij} \right) \ .
\end{eqnarray}
 Using the slow-roll parameters and the following approximate scale factor
\begin{gather}
a(\tau) \simeq -\dfrac{1}{H\tau} \ ,
\end{gather}
we can write the total quadaratic  action for tensor perturbations at the second order as
\begin{align}
\delta S_{\text{tensor}} &\equiv \delta S_{\text{EH}}+\delta S_{\text{scalar}}+\delta S_{\text{gauge}}+\delta S_{\text{CS}} \notag \\
 &= \int d^3\bm{x}d\tau \dfrac{1}{2}\left[ \dfrac{1}{4}\psi'^{ij}\psi'_{ij}-\dfrac{1}{4}\psi^{ij,k}\psi_{ij,k} + \dfrac{1}{2\tau^2}\psi^{ij}\psi_{ij} \right] \notag \\
 &+ \int d^3\bm{x}d\tau \left[  \dfrac{1}{2}t'^a_i t'^a_i - \dfrac{1}{2}t^a_{i,j}t^a_{i,j} + \dfrac{1}{2}\left( \dfrac{I''}{I} - \dfrac{2m_{Q}\xi}{\tau^2} \right)t^a_i t^a_i - \dfrac{1}{\tau}\left( m_{Q} + \xi \right)\epsilon^{ijk}t^a_i t^a_{k,j} \right] \notag \\
 & + \int d^3\bm{x}d\tau \left[ \dfrac{\sqrt{\epsilon_E}}{\tau}\psi^{ij}\left( t'_{ij} - \dfrac{I'}{I}t_{ij} \right) - \dfrac{\sqrt{\epsilon_B}}{\tau}\psi^{jm}\epsilon^a_{ij}t^a_{m,i} + \dfrac{\sqrt{\epsilon_B}m_{Q}}{\tau^2}\psi^{ij}t_{ij} + \dfrac{\epsilon_E - \epsilon_B}{4\tau^2}\psi^{ij}\psi_{ij} \right] \label{eq: tensoraction} \ ,
\end{align}
where we defined a new variable
\begin{equation}
\xi \equiv \dfrac{\lambda}{2I^2}\dfrac{\dot{\bar{\sigma}}}{f H} \ .
\end{equation}
We shall use the interaction picture. We treat
 ~$\psi_{ij}$ ~and ~$t_{ij}$ ~as free fields in the de-Sitter background at the leading order. 
 We can regard the last line in the action \eqref{eq: tensoraction} as the interaction part.
 In order to calculate the power spectrum of gravitational waves, we quantize these variables.
 The canonical quantization gives rise to the following expansion
\begin{align}
\psi_{ij}(\bm{x},\tau) 
&= 2\sum_{A = \pm}\int\dfrac{d^3\bm{k}}{(2\pi)^3} \hat{\psi}^A_{k}(\tau)e^{i\bm{k}\cdot\bm{x}} \ , \\
&= 2\sum_{A = \pm}\int\dfrac{d^3\bm{k}}{(2\pi)^3}\left[e^{A}_{ij}(\hat{\bm{k}})\psi^A_{k}(\tau)a^A_{\bm{k}} + e^{A*}_{ij}(-\hat{\bm{k}})\psi^{A*}_{k}(\tau)a^{A\dagger}_{-\bm{k}} \right]e^{i\bm{k}\cdot\bm{x}} \ , \\
t_{ij}(\bm{x},\tau)
&= 2\sum_{A = \pm}\int\dfrac{d^3\bm{k}}{(2\pi)^3} \hat{t}^A_{k}(\tau)e^{i\bm{k}\cdot\bm{x}} \ , \\
&= \sum_{A = \pm}\int\dfrac{d^3\bm{k}}{(2\pi)^3}\left[e^{A}_{ij}(\hat{\bm{k}})t^A_{k}(\tau)b^A_{\bm{k}} + e^{A*}_{ij}(-\hat{\bm{k}})t^{A*}_{k}(\tau)b^{A\dagger}_{-\bm{k}} \right]e^{i\bm{k}\cdot\bm{x}} \ ,
\end{align}
where $e^{A}_{ij}(\hat{\bm{k}})$ are the polarization tensors which satisfy the following normalization relation, $e^{Aij}(\hat{\bm{k}})e^{B}_{ij}(-\hat{\bm{k}})=\delta^{AB}$, and the index ``$A = \set {+, -}$" represents a circular polarization state defined by $ik^i\epsilon^a_{ij}e^\pm_{jm}(\hat{\bm{k}})=\pm ke^\pm_{am}(\hat{\bm{k}})$ .
 The creation and annihilation operators ~$a^A_{\bm{k}} \ , b^A_{\bm{k}}$ ~satisfy the following commutation relations 
\begin{equation}
[a^A_{\bm{k}} \ , a^{B\dagger}_{-\bm{k}'}] = [b^A_{\bm{k}} \ , b^{B\dagger}_{-\bm{k}'}] = (2\pi)^3\delta_{AB}\delta^3(\bm{k} + \bm{k}') \ , \qquad [a^A_{\bm{k}} \ , b^{B}_{\bm{k}'}] = [a^A_{\bm{k}} \ , b^{B\dagger}_{-\bm{k}'}] = 0 \label{eq: commute} \ .
\end{equation}
 Moreover, ~$\psi^A_{k}$ ~and ~$t^A_{k}$ ~are  mode functions which satisfy the following equations of motion in the de Sitter spacetime
\begin{gather}
\dfrac{d^2\psi^\pm_{k}}{dx^2}+\left( 1-\dfrac{2}{x^2} \right)\psi^\pm_{k} = 0 \label{eq: metrictensor2} \ , \\
\dfrac{d^2 t^\pm_{k}}{dx^2} + \left(1 - \dfrac{d^2 I/dx^2}{I} + \dfrac{2 m_{Q}\xi}{x^2} \mp\dfrac{2(m_{Q}+\xi)}{x} \right)t^\pm_{k} = 0 \label{eq: gaugetensor2} \ .
\end{gather}
 Here we used a dimensionless time variable ~$x\equiv -k\tau$.
 In the in-in formalism \cite{Weinberg:2005vy}, the  in-state is given by
\begin{equation}
\ket{in} \equiv T\exp\left( -i\int^\tau_{-\infty(1 + \epsilon)}d\tilde{\tau} H_I(\tilde{\tau}) \right) \ket{0} \ ,
\end{equation}
where $T$ represents the time-ordered product  and ~$\ket{0}$  is a  vacuum state defined by
\begin{equation}
a^A_{\bm{k}}\ket{0} = b^A_{\bm{k}}\ket{0} = 0 \ , \qquad \braket{0|0} = 1 \ .
\end{equation}
 The interaction Hamiltonian can be read off from the action \eqref{eq: tensoraction} as
\begin{align}
H_I(\tau) &= -\int d^3\bm{x}\left[ \dfrac{\sqrt{\epsilon_E}}{\tau}\psi^{ij}v_{ij} - \dfrac{\sqrt{\epsilon_B}}{\tau}\psi^{jm}\epsilon^a_{ij}t^a_{m,i} + \dfrac{\sqrt{\epsilon_B}m_{Q}}{\tau^2}\psi^{ij}t_{ij} + \dfrac{\epsilon_E - \epsilon_B}{4\tau^2}\psi^{ij}\psi_{ij} \right] \notag \\
&= -2\sum_{A = \pm}\int \dfrac{d^3\bm{k}}{(2\pi)^3}\left[ J^A_{\bm{k}} ~\psi^A_{-\bm{k}} + \dfrac{\epsilon_E - \epsilon_B}{2\tau^2} \psi^A_{\bm{k}} ~\psi^A_{-\bm{k}} \right]  \ ,
\end{align}
where we used new variables
\begin{gather}
v_{ij} \equiv t'_{ij} - \dfrac{I'}{I}t_{ij} \ , \qquad v^A_{\bm{k}} = t^A_{\bm{k}}{'} - \dfrac{I'}{I}t^A_{\bm{k}} \ , \\
J^\pm_{\bm{k}}(\tau) \equiv \dfrac{\sqrt{\epsilon_E}}{\tau}v^\pm_{\bm{k}} + \left( \dfrac{\sqrt{\epsilon_B}m_{Q}}{\tau^2} \pm \dfrac{k\sqrt{\epsilon_B}}{\tau} \right)t^\pm_{\bm{k}} \ .
\end{gather}
 The amplitude of a helicity mode of gravitational waves in the in-state is given by
\begin{align}
&(2\pi)^3\delta(\bm{k}+\bm{k}')\bra{in}h^A_k(\tau)^2\ket{in} =\sum_{N=0}^{\infty}(-i)^N\int^{\tau}d\tau_1 \int^{\tau_1}d\tau_2 \ ... \ \int^{\tau_{N-1}}d\tau_N \notag \\
&\times\bra{0}\left[\left[\left[ h^A_{\bm{k}}(\tau)h^A_{\bm{k'}}(\tau) \ , \ H_I(\tau_1) \right] \ , H_I(\tau_2) \right] \ ... \ , H_I(\tau_N) \right] \ket{0} \label{eq: In-In} \ .
\end{align}
Expanding \eqref{eq: In-In} up to the second order in ~$H_I$ , we have
\begin{align}
\eqref{eq: In-In} &= \dfrac{4}{a(\tau)^2}\left( \bra{0}\psi^A_{\bm{k}}(\tau)\psi^A_{\bm{k'}}(\tau)\ket{0} \right. \label{eq: zero} \\
 &\left. -i\int^{\tau}d\tau_1 \bra{0}\left[ \psi^A_{\bm{k}}(\tau)\psi^A_{\bm{k'}}(\tau) \ , \ H_I(\tau_1) \right]\ket{0} \right. \label{eq: 1st} \\
 &\left. -\int^{\tau}d\tau_1 \int^{\tau_1}d\tau_2 \bra{0}\left[\left[ \psi^A_{\bm{k}}(\tau)\psi^A_{\bm{k'}}(\tau) \ , \ H_I(\tau_1) \right] \ , H_I(\tau_2) \right]\ket{0} \right) \label{eq: 2nd} \\
 &+ (\text{higher order}) \ .
\end{align}
 Using commutation relations \eqref{eq: commute} and normalization conditions for the polarization tensors, we get
\begin{align}
\eqref{eq: 1st} &= (2\pi)^3\delta(\bm{k} + \bm{k}') \times 16H^2\tau^2 ~\text{Im}\left[ \int^\tau d\tau_1 \dfrac{\epsilon_E - \epsilon_B}{\tau_1^2}\psi^A_{k}(\tau_1)^2 \psi^{A*}_{k}(\tau)^2 \right] \ , \label{eq: first} \\
\eqref{eq: 2nd} &= (2\pi)^3\delta(\bm{k} + \bm{k}') \times(-32H^2\tau^2) \int^\tau d\tau_1 \int^{\tau_1} d\tau_2 \notag \\
&\left[ G_k(\tau, \tau_1)G_k(\tau, \tau_2)J^{A*}_k(\tau_1)J^{A}_k(\tau_2) +G_k(\tau, \tau_1)F_k(\tau_1, \tau_2)\psi_k(\tau)\psi^*_k(\tau_2) \right] \ , \label{eq: second}
\end{align}
where we defined the following functions
\begin{align}
G_k(\tau, \tau_i) &\equiv 2i ~\text{Im}\left[ \psi_k(\tau)\psi^*_k(\tau_i) \right] \ , \\
F_k(\tau_1, \tau_2) &\equiv 2i ~\text{Im}\left[ J^{A}_k(\tau_1)J^{A*}_k(\tau_2) \right]
\end{align}
which came from the commutation relations of ~$\hat{\psi}^A_{\bm{k}}$ ~and ~$\hat{J}^A_{\bm{k}}$ .

 As to the mode functions for the metric, we can take the Bunch-Davis (BD) mode functions 
\begin{gather}
\psi^\pm_{k} = \dfrac{1}{\sqrt{2k}}\left( 1 + \dfrac{i}{x} \right)e^{ix} \label{eq: BDmetric} \ .
\end{gather}
 For the gauge field, it is difficult to give an analytic solution for ~$t^A_k$ because $I(\bar{\varphi}) \ , m_{Q}$ ~and ~$\xi$ ~in $\eqref{eq: gaugetensor2}$ are not known analytically.
 Therefore, we approximately solve $t^A_k$ ~in each inflationary stage as discussed in Sec.\ref{ba} , 
and estimate the GW spectrum analytically in each stage.

\subsection{Tensor spectrum in the early stage of inflation}

 Let us estimate the GW spectrum in the early stage of inflation.
 During this stage, ~$m_{Q}$ ~and ~$\xi$ ~are negligible so that the equation of motion for ~$t^\pm_{k}$ ~\eqref{eq: gaugetensor2} 
is approximately given  by
\begin{gather}
\dfrac{d^2 t^\pm_{k}}{dx^2} + \left(1 - \dfrac{d^2 I/dx^2}{I} \right)t^\pm_{k} \simeq 0 \label{eq: gaugetensor3} \ .
\end{gather}
 The time evolution of ~$I(\bar{\varphi})$  depends on ~$n$ ~and an initial value of the gauge field.
 If the gauge field grows sufficiently, however, we obtain the transient attractor ~$I \propto a^{-2}$ as we showed in \eqref{eq: Ia} .
 Therefore, we can approximately calculate as
\begin{equation}
\dfrac{d I/dx}{I} \sim \dfrac{2}{x} \ , \quad \dfrac{d^2 I/dx^2}{I} \sim \dfrac{2}{x^2} \ .
\end{equation}
 Thus, the mode fluctuations for the gauge field in this stage are 
\begin{gather}
t^\pm_{k} \simeq \dfrac{1}{\sqrt{2k}}\left( 1 + \dfrac{i}{x} \right)e^{ix} \label{eq: BD} \ .
\end{gather}
 Substituting \eqref{eq: BDmetric} and \eqref{eq: BD} into \eqref{eq: In-In}, we can evaluate the power spectrum.
 Here, we use the following cosine and sine integrals 
\begin{align}
\text{Ci}(x) &\equiv -\int_x^{\infty}dx\dfrac{\cos x}{x} = \gamma + \ln x + \mathcal{O}(x^2) \ , \\
\text{Si}(x) &\equiv \int_0^x dx\dfrac{\sin x}{x} = x + \mathcal{O}(x^3) \ ,
\end{align}
where $\gamma$ is an Euler number.
 Then, we get
\begin{align}
\eqref{eq: first} \times (16H^2\tau^2)^{-1} 
&= \dfrac{\epsilon_E - \epsilon_B}{3}\left[ \dfrac{1}{x^2} - \dfrac{1}{x}\left( \text{Ci}(2 x)\sin 2 x - \text{Si}(2 x) \cos 2 x + \dfrac{\pi}{2}\cos 2 x \right) \right. \notag \\
&\left. \qquad + \dfrac{x^2 - 1}{2x^2}\left( \text{Ci}(2 x)\cos 2 x + \text{Si}(2 x)\sin 2 x - \dfrac{\pi}{2} \sin 2 x \right) \right] \ ,
\end{align}
and
\begin{align}
&\eqref{eq: second} \times (-32H^2\tau^2)^{-1} \notag \\
&= -\dfrac{\pi^2}{32}\left[ \epsilon_E - \dfrac{4m_{Q}}{3}\sqrt{\epsilon_E\epsilon_B} + \left( 1 + \dfrac{4 m_{Q}^2}{9} \right)\epsilon_B \right]\left(\dfrac{x^2+1}{x^2}\right) - \left( \dfrac{\sqrt{\epsilon_E\epsilon_B}}{4} - \dfrac{m_{Q} \epsilon_B}{12} \right)\dfrac{s_A}{x} \notag \\
& -\left[ \dfrac{13}{12}\epsilon_E - \dfrac{2 m_{Q}}{3}\sqrt{\epsilon_E\epsilon_B} - \left( \dfrac{1}{16} - \dfrac{
 2m_{Q}^2}{27} \right)\epsilon_B \right]\dfrac{1}{x^2} + \left[ \dfrac{11}{12x}\epsilon_E
 +\left(\dfrac{s_A}{4} - \dfrac{2m_{Q}}{3 x} - \dfrac{3 s_A}{8 x^2} \right)\sqrt{\epsilon_E\epsilon_B} \right. \notag \\
 &\left. - \left( \dfrac{s_A m_{Q}}{12} + \dfrac{1}{8 x} - \dfrac{5 m_{Q}^2}{54 x} - \dfrac{s_A m_{Q}}{6 x^2}\right)\epsilon_B \right]\pi \cos 2 x 
- \left[ \left( \dfrac{2}{3} - \dfrac{7}{6 x^2} \right)\epsilon_E - \left( \dfrac{1}{3} m_{Q} + \dfrac{
 5 s_A}{4 x} - \dfrac{m_{Q}}{x^2} \right)\sqrt{\epsilon_E\epsilon_B} \right. \notag \\
&\left. - \left( \dfrac{1}{4} + \dfrac{m_{Q}^2}{54} - \dfrac{s_Am_{Q}}{2 x} + \dfrac{11 m_{Q}^2}{54 x^2} \right)\epsilon_B \right] \left[ \text{Ci}(2 x)\cos 2 x + \text{Si}(2 x)\sin 2 x - \frac{1}{2}\pi \sin 2 x \right] \notag \\
& - \left( \dfrac{1}{8} \epsilon_E - \dfrac{1}{6} m_{Q}\sqrt{\epsilon_E\epsilon_B} + \dfrac{1}{8} \epsilon_B + \dfrac{1}{18} m_{Q}^2\epsilon_B \right)\left(\dfrac{x^2+1}{x^2}\right)\left[\text{Ci}(2 x)^2 + \text{Si}(2x)^2 - \pi\text{Si}(2x)\right] \notag \\
&+\left[ \frac{11}{6 x}\epsilon_E + \left( \frac{s_A}{2} - \frac{4 m_{Q}}{3x} - \frac{3
s_A}{4x^2} \right)\sqrt{\epsilon_E\epsilon_B} -\left( \frac{s_Am_{Q}}{6} + \frac{1}{4 x} - \frac{5m_{Q}^2}{27 x} - \frac{s_Am_{Q}}{3 x^2} \right)\epsilon_B \right]\left[\text{Ci}(2 x) \sin 2 x - \text{Si}(2x) \cos 2 x \right] \notag \\
&+\text{Re}\left[ \left( \dfrac{\epsilon_E}{4} - \left( \dfrac{m_{Q}}{6} + i\dfrac{s_A}{2}\right)\sqrt{\epsilon_E\epsilon_B} - \left(\dfrac{1}{4} - i\dfrac{s_Am_{Q}}{6}\right)\epsilon_B \right)\left( 1 - \dfrac{2i}{x} - \dfrac{1}{x^2} \right)e^{-2ix}\int^x dx_1 \dfrac{1}{x_1}\left( \text{Ci}(2x_1) + i\text{Si}(2x_1) - i\dfrac{\pi}{2} \right) \right] \ .
\end{align}
 Note that we omitted ~$(2\pi)^3\delta(\bm{k} + \bm{k}')$ ~in the above results.
 Moreover, we approximately treated the slow-roll parameters as constants 
 since their time dependence are suppressed by the slow-roll parameters.
 In the super-horizon limit ~$x \rightarrow 0$ , we have
\begin{equation}
\mathcal{P}^{\pm}_h(k) = \bra{in}h^\pm_k(\tau)^2\ket{in} \simeq \dfrac{2H^2}{k^3}\left[ 1 + 4 \left(\epsilon_E - m_{Q}\sqrt{\epsilon_E\epsilon_B} + \dfrac{2}{9}m_{Q}^2\epsilon_B \right)(\ln x)^2 \right] \label{eq: gwCMB} \ .
\end{equation}
 The power spectrum of the gravitational waves is not constant in the super-horizon  due to the interaction with the gauge field.
 However, its modification will not be so serious unless the duration of this period is too large.
 This tensor spectrum in this stage is generated around CMB scales and we can see no parity-violation in the spectrum.

\subsection{Chiral gravitational waves in the late stage of inflation}

 Now, we calculate the spectrum in the late stage of inflation.
 In this stage,  ~$\xi$ ~and ~$m_Q$ ~contribute the background dynamics.
 From the slow-roll equation \eqref{eq: barxi}, we have
\begin{equation}
\xi \simeq m_{Q} + \dfrac{1}{m_{Q}}\left( 1 + \dfrac{\dot{I}}{HI} \right) \ .
\end{equation}
 As to the gauge kinetic function, it evolves as ~$x^{-nV/3H^2} \ (\because \eqref{eq: Ib})$. Hence,we can calculate as
\begin{gather}
\dfrac{d I/dx}{I} \sim -\dfrac{\bm{n}}{x} \ , \quad \dfrac{d^2 I/dx^2}{I} \sim \dfrac{\bm{n}(\bm{n}+1)}{x^2} \ , \quad \bm{n} \equiv nV/3H^2 \lesssim 1 \ .
\end{gather}
 Note that the energy density of the axion is greater than that of the dilaton in this stage.
 Then, Eq.\eqref{eq: gaugetensor2} reads
\begin{eqnarray}
\dfrac{d^2 t^\pm_{k}}{dx^2} + \left(1+\dfrac{A}{x^2} \mp \dfrac{2B}{x} \right)t^\pm_{k} \simeq 0 \ , 
\end{eqnarray}
where
\begin{eqnarray}
A \equiv 2\left( m_{Q}^2 + 1 - \dfrac{\bm{n}(\bm{n} - 1)}{2} \right) \ , \qquad B \equiv 2m_{Q} + \dfrac{1}{m_{Q}}\left( 1 + \bm{n} \right) \ .
\end{eqnarray}
 We can neglect time dependence of $A$ and $B$.
 Here, we consider the parameter region where both $A$ and $B$ have positive values.
 We can see that the mass term of ~$t^-_k$ ~is always positive, so this mode is stable.
 On the other hand, $t^+_k$  has the tachyonic instability  during the time interval
\begin{equation}
\dfrac{1}{2}(B - \sqrt{B^2 - A}) < x < \dfrac{1}{2}(B + \sqrt{B^2 - A})
\end{equation}
and this instability can enhance one of  helicity modes of gravitational waves.

 Let us focus on the $t^+_k$ mode.
 It is known that  solutions are given by Whittaker functions
\begin{equation}
t^+_{k}(x) = \dfrac{1}{\sqrt{2k}}\left(C_1 M_{\kappa, \mu}(2ix) + C_2 W_{\kappa, \mu}(2ix)\right) \label{eq: gaugeWhi} \ ,
\end{equation}
where $C_1$ and $C_2$ are constants and
\begin{equation}
\kappa \equiv i B \ , \qquad \mu^2 \equiv \dfrac{1}{4} - A \ .
\end{equation}
 In the sub-horizon limit $x \rightarrow \infty$ , these functions have asymptotic expansion 
\begin{align}
M_{\kappa, \mu}(2ix) &\simeq \dfrac{\Gamma(1+2\mu)}{\Gamma(\frac{1}{2} + \mu - \kappa)}(2i)^{-\kappa}e^{i(x + i\kappa\ln x)} + \dfrac{\Gamma(1+2\mu)}{\Gamma(\frac{1}{2} + \mu + \kappa)} (-1)^{\frac{1}{2} + \mu - \kappa} (2i)^{\kappa} e^{-i(x + i\kappa\ln x)} \ , \\
W_{\kappa, \mu}(2ix) &\simeq (2i)^\kappa e^{-i(x + i\kappa\ln x)} \ .
\end{align}
In order for the mode function to describe the BD vacuum, the constants $C_1$ ~and ~$C_2$ should be 
\begin{equation}
C_1 = \dfrac{\Gamma(\frac{1}{2} + \mu - \kappa)}{\Gamma(1 + 2\mu)}(2i)^\kappa \ , \qquad C_2 = -\dfrac{\Gamma(\frac{1}{2} + \mu - \kappa)}{\Gamma(\frac{1}{2} + \mu + \kappa)}(2i)^\kappa (-1)^{\frac{1}{2} + \mu - \kappa} \ .
\end{equation}

\begin{figure}[h]
\begin{center}
\includegraphics[width=8cm,height=8cm,keepaspectratio]{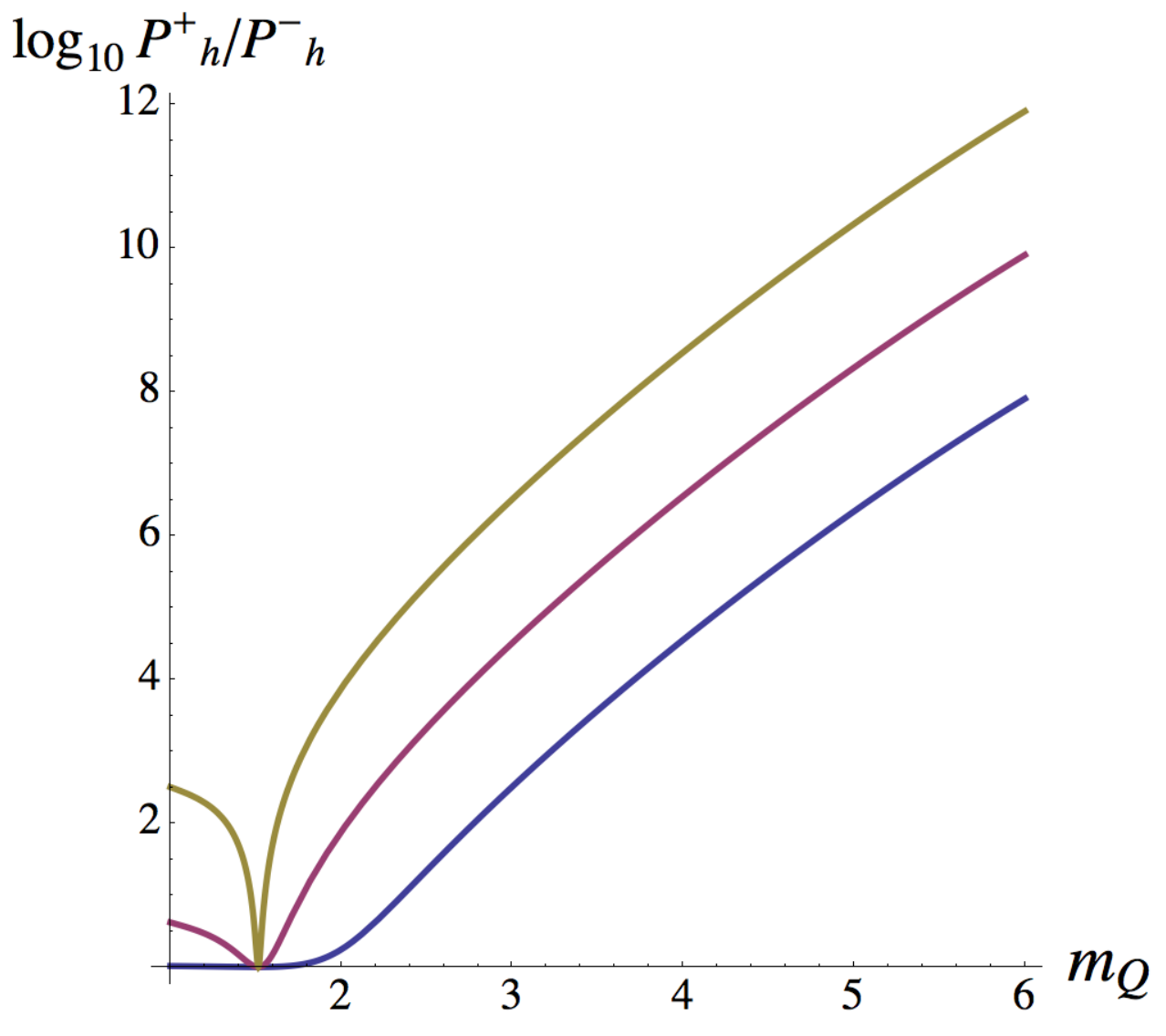}
\end{center}
\caption{The behaviour of the ratio of the GW spectra as a function of  ~$m_Q$ ~for ~$Q_{\text{min}}=10^{-2}$ ~(blue line) , ~$Q_{\text{min}}=10^{-1}$ ~(red line) and ~$Q_{\text{min}}=1$ ~(yellow line) .}
\label{fig: CGWratio}
\end{figure}

 Now, we can calculate the power spectrum of gravitational waves in this stage.
 The dominant contribution comes from an integration \eqref{eq: second}
 where the gauge field fluctuations experience the tachyonic instability near the horizon crossing.
 In this region, Whittaker M-function is a decaying mode and irrelevant for the integration.
 Moreover, during this stage we can regard the complex phase of $t^+_{k}$ as nearly constant 
due to its enhancement and approximately treat ~$t^+_{k}$ ~as a classical variable\footnote{
The detail is  discussed in \cite{Barnaby:2011qe, Barnaby:2012tk}}.
 As a result, the contribution from ~$C_1 M_{\kappa, \mu}(2ix)$ ~and the commutators including ~$J^+_{\bm{k}}$ ~will be numerically negligible.
 Hence, we  can use
\begin{gather}
t^+_{k}(x) \simeq C_2 W_{\kappa, \mu}(2ix) \ , \\
F_k(\tau_1, \tau_2) \simeq 0
\end{gather}
in the integration \eqref{eq: second} .
 Then we can evaluate the integration as
\begin{align}
\eqref{eq: second} &\simeq (2\pi)^3\delta(\bm{k} + \bm{k}') \times(16H^2\tau^2) \left| \int^\tau d\tau_1 G_k(\tau, \tau_1)J^{A}_k(\tau_1) \right|^2 \ . \label{eq: second2}
\end{align}
 We can perform this integration by using the following identities
\begin{align}
\int dx x^ne^{ix}W_{\kappa, \mu}(2i x) &=  \dfrac{x^{n+1}G^{2,2}_{2,3}\left( 2i x \left|
\begin{array}{ccc}
-n, & 1+\kappa  \\
 \frac{1}{2} - \mu, & \frac{1}{2} + \mu, & -1 - n 
\end{array}
 \right. \right)}{\Gamma(\frac{1}{2} - \kappa -\mu)\Gamma(\frac{1}{2} - \kappa + \mu)} \ , \\
\int dx x^ne^{-ix}W_{\kappa, \mu}(2i x) &= x^{n+1}G^{2,1}_{2,3}\left( 2i x \left|
\begin{array}{ccc}
-n, & 1-\kappa  \\
 \frac{1}{2} - \mu, & \frac{1}{2} + \mu, & -1 - n
\end{array}
 \right. \right) \ ,
\end{align}
where G is the Meijer G-function.
Thus, we get the GW power spectrum. For the stable circular polarization state,  we obtain the conventional spectrum
\begin{equation}
\mathcal{P}^{-}_h(k) = \bra{in}h^-_k(\tau)^2\ket{in} \simeq \dfrac{2H^2}{k^3} \label{eq: GW-} \ .
\end{equation}
For the other circular polarization state, due to the enhancement of the gauge field fluctuations, we have
\begin{align}
\mathcal{P}^{+}_h(k) = \bra{in}h^+_k(\tau)^2\ket{in} &= \dfrac{2H^2}{k^3}\left[ 1 + 8|C_2|^2 \left| \sqrt{\epsilon_E}\mathcal{I}_0(m_{Q}) \right. - \sqrt{\epsilon_B}\mathcal{I}_1(m_{Q}) + ( \bm{n}\sqrt{\epsilon_E} \left.+ m_{Q}\sqrt{\epsilon_B})\mathcal{I}_2(m_{Q}) \right|^2 \right] \notag \\
 &\simeq \dfrac{2H^2}{k^3}\left[ 1 + 8 |C_2|^2Q_{\text{min}}^2\left| \mathcal{I}_0(m_{Q}) \right. - m_{Q}\mathcal{I}_1(m_{Q}) \left.+ m_{Q}^2 \mathcal{I}_2(m_{Q}) \right|^2 \right] \label{eq: GW+} \ ,
\end{align}
where we defined
\begin{align}
\mathcal{I}_0(m_{Q}) &= \dfrac{ i~\Gamma(-\frac{3}{2} - \mu)\Gamma(-\frac{3}{2} + \mu)}{2}\left(\dfrac{ (\frac{1}{4} - \mu^2- 4\kappa)(\frac{9}{4} - \mu^2) + 8\kappa(1 + \kappa) }{\Gamma(1-\kappa)} \right. \notag \\
&\left. - \dfrac{(\frac{1}{4} - \mu^2 + 4\kappa)(\frac{9}{4} - \mu^2) - 8\kappa(1 - \kappa)}{\Gamma(\frac{1}{2}-\mu-\kappa)\Gamma(\frac{1}{2}+\mu-\kappa)\Gamma(-\kappa)^{-1}} \right) \ , \\
\mathcal{I}_1(m_{Q}) &= \dfrac{\Gamma(-\frac{1}{2} - \mu)\Gamma(-\frac{1}{2} + \mu)}{2}\left( \dfrac{\frac{1}{4} - \mu^2 - 2\kappa}{\Gamma(1-\kappa)} + \dfrac{\frac{1}{4} - \mu^2 + 2\kappa}{\Gamma(\frac{1}{2}-\mu-\kappa)\Gamma(\frac{1}{2}+\mu-\kappa)\Gamma(-\kappa)^{-1}} \right) \ , \\
\mathcal{I}_2(m_{Q}) &= i~\Gamma(-\tfrac{1}{2} - \mu)\Gamma(-\tfrac{1}{2} + \mu)\left( \dfrac{1 - 2(1 + \kappa)(\frac{1}{4} - \mu^2)}{\Gamma(-\kappa)} + \dfrac{1 - 2(1 - \kappa)(\frac{1}{4} - \mu^2)}{\Gamma(\frac{1}{2}-\mu-\kappa)\Gamma(\frac{1}{2}+\mu-\kappa)\Gamma(1-\kappa)^{-1}} \right) \ .
\end{align}
 We plotted the ratio of spectra \eqref{eq: GW-} and \eqref{eq: GW+} in FIG. \ref{fig: CGWratio} .
 We can see that chiral gravitational waves are more enhanced as ~$m_Q$ ~is increasing.
 This qualitative feature is the same as that of chromo-natural inflation \cite{Adshead:2013nka}. 

 Remarkably, this model can avoid the overproduction of chiral gravitational waves on CMB scales discussed in previous works \cite{Dimastrogiovanni:2012ew, Adshead:2013nka} and produce a sizable circulary polarized gravitational waves on small scales.
 In the next section, we also study the dynamics of scalar perturbations and show that this model is stable under the scalar fluctuations
 and consistent with the CMB data.

\section{Viability of the model \label{sca}}

 In this section, we check the dynamics of scalar perturbations.
 Specifically, we compute the curvature perturbation and its spectrum in the early stage of inflation where fluctuations on  CMB scales are created.
 Moreover, we discuss the stability of scalar perturbations in the late stage of inflation.

 Let us derive  the quadratic action for scalar perturbations.
 The Einstein-Hilbert action gives
\begin{equation}
\delta S_{\text{EH}}= \int d\tau d\bm{x} \left[ -12a^4H^2 \phi^2 - 4a^2H\phi\partial^2\beta \right] \ ,
\end{equation}
where ~$\phi$ ~and ~$\beta$ ~are perturbed lapse and shift functions defined in \eqref{eq: metper} .
 The (pseudo) scalar quadratic actions  are given by
\begin{align}
\delta S_{\text{dilaton}} &= \int d\tau d\bm{x} \left[ 2a^4\dot{\bar{\varphi}}^2\phi^2 - 2\phi(a^3\dot{\bar{\varphi}} \delta \varphi' + a^4V_{\bar{\varphi}}\delta\varphi) + a^2\dot{\bar{\varphi}}\beta\partial^2\delta\varphi \right. \notag \\
 &\left. + \dfrac{1}{2}a^2\delta \varphi'^2 + \dfrac{1}{2}a^2\delta \varphi\partial^2\delta\varphi - \dfrac{1}{2}a^4V_{\bar{\varphi}\bar{\varphi}}\delta \varphi^2  \right] \ , \notag \\
\delta S_{\text{axion}} &= \int d\tau d\bm{x} \left[ 2a^4\dot{\bar{\sigma}}^2\phi^2 - 2\phi(a^3\dot{\bar{\sigma}} \delta \sigma' + a^4W_{\bar{\sigma}}\delta\sigma) + a^2\dot{\bar{\sigma}}\beta\partial^2\delta\sigma \right. \notag \\
 &\left. + \dfrac{1}{2}a^2\delta \sigma'^2 + \dfrac{1}{2}a^2\delta \sigma\partial^2\delta\sigma - \dfrac{1}{2}a^4W_{\bar{\sigma}\bar{\sigma}}\delta \sigma^2  \right] \ .
\end{align}
 For the quadratic action $\delta S_{\text{gauge}}$ , we split it into two parts ~$\delta S^{\text{I}}_{\text{gauge}} + ~\delta S^{\text{I\hspace{-.1em}I}}_{\text{gauge}}$ : the former one includes only gauge field fluctuations 
\begin{align}
\delta S^{\text{I}}_{\text{gauge}} &= \int d\tau d\bm{x} I^2\left[ \dfrac{3}{2}(a\delta Q)'^2 - (a U)'\partial^2(a U)' +\dfrac{1}{2}a^4Y\partial^4Y - a^6g^2Q^2Y\partial^2Y  \right. \notag \\
 &\left. - a^2Y\left( \partial^2(a\delta Q)' + 2agQ\partial^2(aU)' - 2a^3\dfrac{\dot{(aQ)}}{a}g\partial^2U \right) \right. \notag \\
 &\left. + a^2\delta Q\partial^2\delta Q - a^2\delta U\partial^4\delta U - 6a^3gQ\delta Q\partial^2U - a^4g^2Q^2\left(9\delta Q^2 - 2U\partial^2U \right) \right] \label{eq: gau1per} \ ,
\end{align}
 and the latter one has the interaction with other fields 
\begin{align}
\delta S^{\text{I\hspace{-.1em}I}}_{\text{gauge}} &= \int d\tau d\bm{x} \left[ 4a^4\rho_E\phi^2 - \dfrac{2}{3}a^2\rho_B\beta\partial^2\beta - 4a^4\phi\dfrac{I_{\bar{\varphi}}}{I}(\rho_E + \rho_B)\delta\varphi \right. \notag \\
 &\left. +2\beta a I^2\left(\dfrac{\dot{(aQ)}}{a}a\partial^2\delta Q + gQ^2(a\partial^2U)' + a^3g^2Q^3\partial^2 Y - \dfrac{\dot{(aQ)}}{a}a^2gQ\partial^2U \right) \right. \notag \\
 &\left. -2a^2I^2(\phi - \dfrac{I_{\bar{\varphi}}}{I}\delta\varphi)\dfrac{\dot{(aQ)}}{a}\left(3(a\delta Q)'-a^2\partial^2Y \right) \right. \notag \\
 &\left. -4a^3I^2(\phi + \dfrac{I_{\bar{\varphi}}}{I}\delta\varphi)gQ^2\left(3agQ\delta Q +\partial^2U \right) \right. \notag \\
 &\left. + \left( \dfrac{I_{\bar{\varphi}\bar{\varphi}}}{I} + \dfrac{I_{\bar{\varphi}}^2}{I^2} \right)(\rho_E - \rho_B)a^4\delta\varphi^2  \right] \label{eq: gau2per} \ .
\end{align}
 Finally, the quadratic action $\delta S_{\text{CS}}$ ~reads
\begin{align}
\delta S_{\text{CS}} &= \int a d\tau d\bm{x} \left[ \dfrac{\lambda\dot{\sigma}}{f}\left( 2\delta Q \partial^2 U + agQ(3\delta Q^2 - U\partial^2 U) \right) \right. \notag \\
 &\left.+\dfrac{\lambda}{f}\left( Ya^2gQ^2\partial^2\delta\sigma + 6agQ^2\delta\sigma'\delta Q - 2\dot{(aQ)}\delta\sigma\partial^2U \right) \right] \label{eq: CSper} \ .
\end{align}

\subsection{Curvature perturbation spectrum in the early stage of inflation \label{cur}}

 The curvature perturbations in the uniform density gauge is defined by \cite{Malik:2008im}
\begin{equation}
-\zeta \equiv \psi + \dfrac{H}{\dot{\bar{\rho}}}\delta\rho \ ,
\end{equation}
where we defined the energy density of matter fields as $\rho = \bar{\rho} + \delta\rho$.
 The energy momentum tensor of matter  is given by
\begin{align}
T_{\mu\nu} &= -\dfrac{2}{\sqrt{-g}}\dfrac{\delta S_{\text{matter}}}{\delta g^{\mu\nu}} \\
 &\equiv (\rho + P)u_\mu u_\nu + P g_{\mu\nu} \ ,
\end{align}
where ~$P$ ~is the pressure of matter and ~$u^{\mu} = \frac{dx^\mu}{d\tau}$ ~is the four-velocity which satisfies
\begin{equation}
u^{\mu}u_{\mu} = -1 \ .
\end{equation}
 Therefore, the background energy density reads
\begin{align}
\bar{\rho} &= -\bar{T}^0_0 \notag \\
 &= \dfrac{1}{2}\dot{\bar{\varphi}}^2 + V(\bar{\varphi}) + \dfrac{1}{2}\dot{\bar{\sigma}}^2 + W(\bar{\sigma}) + \rho_E + \rho_B \ .
\end{align}
and its fluctuation at the first order is given by
\begin{align}
\delta\rho &= -\delta T^0_0 \\
 &\equiv \delta\rho_{\text{dilaton}} + \delta\rho_{\text{axion}} + \delta\rho_{\text{gauge}} \ ,
\end{align}
where we split $\delta\rho$ into three variables in the second line, which are derived from each action $S_{\text{dilaton}} \ , \ S_{\text{axion}}$ ~and ~$S_{\text{gauge}}$ . 
 More concretely, we have
\begin{align}
\delta\rho_{\text{dilaton}} &= \dfrac{1}{a^2}\bar{\varphi}' \delta\varphi' + V_{\bar{\varphi}}\delta\varphi - \dfrac{2}{a^2}\bar{\varphi}'^2\phi \ , \\
\delta\rho_{\text{axion}} &= \dfrac{1}{a^2}\bar{\sigma}' \delta\sigma' + W_{\bar{\sigma}}\delta\sigma - \dfrac{2}{a^2}\bar{\sigma}'^2\phi
\end{align}
and
\begin{equation}
\delta\rho_{\text{gauge}} = 6I^2\dfrac{\dot{(aQ)}^2}{a^2}\phi + 3\left(\dfrac{\dot{(aQ)}^2}{a^2} + g^2Q^4\right)I I_{\bar{\varphi}}\delta\varphi + I^2 \dfrac{\dot{(aQ)}}{a} \left( 3 \dfrac{\dot{(a\delta Q)}}{a} - \partial^2 Y \right) + \dfrac{2}{a}I^2g Q^2\partial^2 U + 6I^2g^2 Q^3\delta Q \ .
\end{equation}

 In the early stage of inflation, the background effect of the magnetic field is negligibly small.
 So we can set ~$g=0$ ~and estimate the power spectrum of the curvature perturbation.
 We can easily eliminate $\beta$  and solve $\phi$ as
\begin{equation}
\phi = \dfrac{1}{2}\left( \dfrac{\dot{\bar{\varphi}}}{H}\delta\varphi + \dfrac{\dot{\bar{\sigma}}}{H}\delta\sigma + 2\dfrac{\dot{(aQ)}}{aH}I^2\delta Q \right) \ .
\end{equation}
 Eliminating $Y$  and defining the following variables
\begin{align}
\delta\varphi &= \dfrac{\Delta_\varphi}{a} \ , \quad \delta\sigma = \dfrac{\Delta_\sigma}{a} \ ,  \notag \\
\delta Q &= \dfrac{\Delta_Q}{\sqrt{2}aI} \ , \quad U = \dfrac{\Delta_U}{\sqrt{2}a I} \ ,
\end{align}
we get the following quadratic action for scalar perturbations
\begin{align}
\delta S_{\text{scalar}} &\equiv \delta S_{\text{EH}}+\delta S_{\text{scalar}}+\delta S_{\text{gauge}}+\delta S_{\text{CS}} \notag \\
 &= \int d^3\bm{x}d\tau \dfrac{1}{2}\left[ \Delta_\varphi'^2 + \Delta_\varphi \partial_i^2\Delta_\varphi + \Delta_\sigma'^2 + \Delta_\sigma \partial_i^2\Delta_\sigma + \Delta_Q'^2 + \Delta_Q \partial_i^2\Delta_Q - \Delta_U' \partial_i^2 \Delta_U' - \Delta_U \partial_i^4\Delta_U \right] \notag \\
 & - \int d^3\bm{x}d\tau \dfrac{1}{2}\left[ M_\varphi^2\Delta_\varphi^2 + M_\sigma^2\Delta_\sigma^2 + M_Q^2\Delta_Q^2 - M_U^2\Delta_U\partial_i^2\Delta_U \right] \notag \\
 &+ \int d^3\bm{x}d\tau \left[ -2\frac{\sqrt{2\epsilon _E}}{\tau} \frac{I_{\bar{\varphi}}}{I} \Delta_Q' \Delta_{\varphi} \right] \notag \\
& + \int d^3\bm{x}d\tau \left[ M_{\varphi\sigma}\Delta_\varphi \Delta_\sigma + M_{\varphi Q}\Delta_\varphi \Delta_Q + M_{\sigma Q}\Delta_\sigma \Delta_Q - M_{\sigma U}\Delta_\sigma \partial_i^2 \Delta_U - M_{Q U}\Delta_Q \partial_i^2 \Delta_U \right] \label{eq: scalarpertaction} \ ,
\end{align}
where we defined effective mass parameters
\begin{align}
M_Q^2 &\equiv -\dfrac{2}{\tau^2} +  \dfrac{\epsilon_E}{\tau^2}( 6 - \epsilon_\varphi - \epsilon_\sigma - \epsilon_E ) \ , \\
M_U^2 &\equiv -\dfrac{2}{\tau^2} \ , \\
M_\varphi^2 &\equiv -\frac{2}{\tau^2} + \frac{V_{\bar{\varphi}\bar{\varphi}} + \sqrt{2}
   V_{\bar{\varphi}} \sqrt{\epsilon _{\varphi }} - H^2 \left( \epsilon_{\varphi} + \epsilon_{\sigma}+\epsilon _E \right)\epsilon _{\varphi}}{H^2 \tau^2} - \frac{\epsilon _E}{\tau^2}\left(3 \dfrac{I_{\bar{\varphi}\bar{\varphi}}}{I} - \left(\dfrac{I_{\bar{\varphi}}}{I}\right)^2 - \sqrt{2} \dfrac{I_{\bar{\varphi}}}{I}
   \sqrt{\epsilon _{\varphi}} \right) \ , \\
M_\sigma^2 &\equiv -\frac{2}{\tau^2} + \frac{W_{\bar{\sigma}\bar{\sigma}}+\sqrt{2} W_{\bar{\sigma}} \sqrt{\epsilon _{\sigma }} - H^2\left( \epsilon_{\varphi} + \epsilon_{\sigma}+\epsilon _E \right)\epsilon _{\sigma}}{H^2 \tau^2}
\end{align}
and
\begin{align}
M_{\varphi\sigma} &\equiv -\sqrt{\frac{\epsilon _{\sigma}}{2}}\frac{\epsilon _E I_{\bar{\varphi}}}{\tau^2 I} - \sqrt{\frac{\epsilon _{\sigma}}{2}}\frac{V_{\bar{\varphi}}}{H^2 \tau^2} - \sqrt{\frac{\epsilon_{\varphi}}{2}}\frac{W_{\bar{\sigma}}}{H^2 \tau^2} + \frac{\sqrt{\epsilon_{\varphi}}\sqrt{\epsilon_{\sigma}} \left(\epsilon _{\sigma}+\epsilon _{\varphi}+\epsilon_E \right)}{\tau^2}  \ , \\
M_{\varphi Q} &\equiv \sqrt{\frac{\epsilon_E}{2}}\frac{I_{\bar{\varphi}}}{I} \frac{\left(8 -\epsilon _E \right)}{\tau^2} + \frac{\sqrt{\epsilon_E\epsilon_{\varphi}} \left( - 3 + \epsilon_{\sigma }+\epsilon _{\varphi }+\epsilon _E \right)}{\tau^2} - \sqrt{\frac{\epsilon_E}{2}}\frac{V_{\bar{\varphi}}}{H^2\tau^2} \ , \\
M_{\sigma Q} &\equiv \frac{\sqrt{\epsilon_E\epsilon_{\sigma}} \left( - 3 + \epsilon _{\sigma}+\epsilon _{\varphi}+\epsilon _E \right)}{\tau^2} - \sqrt{\frac{\epsilon_E}{2}}\frac{W_{\bar{\sigma}}}{H^2\tau^2} \ , \\
M_{\sigma U} &\equiv -\frac{\lambda}{I^2 f}\frac{\sqrt{2\epsilon_E}}{\tau} \ , \\
M_{Q U} &\equiv \frac{\xi }{\tau} \ .
\end{align}
 Note that in the above calculations we used the following
\begin{equation}
\dfrac{I'}{I} \sim \dfrac{2}{\tau} \ , \quad \dfrac{I''}{I} \sim \dfrac{2}{\tau^2}
\end{equation}
and assumed $\dot{\bar{\varphi}} \ , \ \dot{\bar{\sigma}} > 0$ so that we can get
 $\frac{\dot{\bar{\varphi}}}{H} = \sqrt{2\epsilon_\varphi}$ ~and ~$\frac{\dot{\bar{\sigma}}}{H} = \sqrt{2\epsilon_\sigma}$ .
 The fluctuation of the energy density is given by
\begin{align}
\delta\rho &= -\dfrac{1}{a^3}\left[ \dfrac{\sqrt{2\epsilon_\varphi}}{\tau}\Delta_\varphi' + \dfrac{\sqrt{2\epsilon_\sigma}}{\tau}\Delta_\sigma'  + \dfrac{\sqrt{2\epsilon_E}}{\tau}\left( \Delta_Q' - \dfrac{2}{\tau}\Delta_Q \right) \right. \notag \\
&\left. + \left( \dfrac{\sqrt{2\epsilon_\varphi}}{\tau^2} - a^2V_{\bar{\varphi}} -  \dfrac{I_{\bar{\varphi}}}{I}\dfrac{\epsilon_E}{\tau^2} \right)\Delta_\varphi + \left( \dfrac{\sqrt{2\epsilon_\sigma}}{\tau^2} - a^2W_{\bar{\sigma}}\right)\Delta_\sigma \right. \notag \\
&\left. + \dfrac{\epsilon_{\varphi}+\epsilon_\sigma - 2\epsilon_E}{\tau^2}\left(\sqrt{2\epsilon_\varphi}\Delta_\varphi+\sqrt{2\epsilon_\sigma}\Delta_\sigma+\sqrt{2\epsilon_E}\Delta_Q \right) \right] \ .
\end{align}

 In order to estimate the curvature perturbation at the leading order, 
let us compare the magnitude of each coefficient (slow-roll parameters) in ~$\delta\rho$.
 From the discussion in Sec.\ref{inea}, ~$\epsilon_\varphi$ ~and ~$\epsilon_E$ ~are dominant in the early stage of inflation
 while ~$\epsilon_\sigma$ ~is negligible.
 Hence ~$\zeta$ ~in the flat slicing gauge is approximately given by
\begin{align}
\zeta &\simeq \dfrac{H\tau^2}{3\sqrt{2\epsilon_\varphi}}\left(1 + \mathcal{E}_E^2 \right)^{-1}\left[ \Delta_\varphi' + \mathcal{E}_E\left( \Delta_Q' - \dfrac{2}{\tau}\Delta_Q \right) + 4\dfrac{1  + \mathcal{E}_E^2}{\tau}\Delta_\varphi \right] \ ,
\end{align}
where we defined a new variable
\begin{equation}
\mathcal{E}_E \equiv \sqrt{\dfrac{\epsilon_E}{\epsilon_\varphi}} \ .
\end{equation}
 We can see that ~$\mathcal{E}_E$ ~can be of order unity.

 Let us calculate the power spectrum of $\zeta$ by using the in-in formalism.
 We canonically quantize the fields as
\begin{align}
\Delta_{\varphi}(\bm{x},\tau) &= \int\dfrac{d^3\bm{k}}{(2\pi)^3}\hat{\Delta}_{\varphi \bm{k}}(\tau)e^{i\bm{k}\cdot\bm{x}} \notag \\
&= \int\dfrac{d^3\bm{k}}{(2\pi)^3}\left[\Delta_{\varphi k}(\tau)c_{\bm{k}} + \Delta^*_{\varphi k}(\tau)c^{\dagger}_{-\bm{k}} \right]e^{i\bm{k}\cdot\bm{x}} \ , \\
\Delta_{Q}(\bm{x},\tau) &= \int\dfrac{d^3\bm{k}}{(2\pi)^3}\hat{\Delta}_{Q \bm{k}}(\tau)e^{i\bm{k}\cdot\bm{x}} \notag \\
&= \int\dfrac{d^3\bm{k}}{(2\pi)^3}\left[\Delta_{Q k}(\tau)d_{\bm{k}} + \Delta^*_{Q k}(\tau)d^{\dagger}_{-\bm{k}} \right]e^{i\bm{k}\cdot\bm{x}} \ ,
\end{align}
where ~$\Delta_{\varphi k}$ ~and ~$\Delta_{Q k}$ ~are  mode functions in de Sitter spacetime.
 From \eqref{eq: scalarpertaction}, we can choose mode functions corresponding the BD vacuum as 
\begin{align}
\Delta_{\varphi k} &\simeq \dfrac{1}{\sqrt{2k}}\left( 1 + \dfrac{i}{x} \right)e^{ix} \ , \\
\Delta_{Q k} &\simeq \dfrac{1}{\sqrt{2k}}\left( 1 + \dfrac{i}{x} \right)e^{ix} \ .
\end{align}
 The creation and annihilation operators ~$c_{\bm{k}} \ , d_{\bm{k}}$  satisfy
\begin{equation}
[c_{\bm{k}} \ , c^{\dagger}_{-\bm{k}'}] = [d_{\bm{k}} \ , d^{\dagger}_{-\bm{k}'}] = (2\pi)^3\delta^3(\bm{k} + \bm{k}') \ , \qquad [c_{\bm{k}} \ , d_{\bm{k}'}] = [c_{\bm{k}} \ , d^{\dagger}_{-\bm{k}'}] = 0 \label{eq: commute2}
\end{equation}
and the vacuum state is defined by
\begin{equation}
c_{\bm{k}}\ket{0} = d_{\bm{k}}\ket{0} = 0 \ , \qquad \braket{0|0} = 1 \ .
\end{equation}
 The interaction Hamiltonian at the leading order reads
\begin{align}
H_I(\tau) &= 4\int d\bm{x}\left[ \dfrac{\mathcal{E}_E^2}{\tau^2}\Delta_\varphi^2 - \frac{\mathcal{E}_E}{\tau}\left( \Delta_Q' - \dfrac{2}{\tau}\Delta_Q \right)\Delta_\varphi \right] \ .
\end{align}
 Thus, we expect that in the super-horizon limit $x \rightarrow 0$, the power spectrum of $\zeta$ takes the following form
\begin{equation}
\mathcal{P}_\zeta(k)  = \bra{in}\zeta_k(\tau)^2\ket{in} \simeq \dfrac{H^2}{4\epsilon_\varphi k^3}\left( 1 + \mathcal{A}~\mathcal{E}_E^2~(\ln x)^2 \right) \ ,
\end{equation}
where ~$\mathcal{A}\sim \mathcal{O}(10)$ ~is a numerical factor.
 Here, note that we neglected the time-derivative of slow-roll parameters in the above estimation.
 This result is consistent with that for  dilatonic inflation coupled to a triad of gauge fields \cite{Yamamoto:2012sq}.
 This power spectrum is also not constant in the super-horizon  due to the interaction with the gauge field.

\subsection{Scalar perturbation stability in the late stage of inflation}

 We are interested in the stability conditions of scalar perturbations in the late stage of inflation
 when the background gauge field settles in the attractor value and chiral gravitational waves are generated.
 For simplicity, we take the background gauge field function ~$Q(t)$ ~to be fixed in the following calculations.
 Moreover, we neglect the perturbations of the metric and set ~$\phi = \beta = 0$ because these contributions are suppressed 
by slow-roll parameters. 
 This truncation is known to be valid from the previous works on the stability of chromo-natural inflation \cite{Dimastrogiovanni:2012ew, Adshead:2013nka}. 

\begin{figure}[h]
\begin{center}
\includegraphics[width=7cm,height=7cm,keepaspectratio]{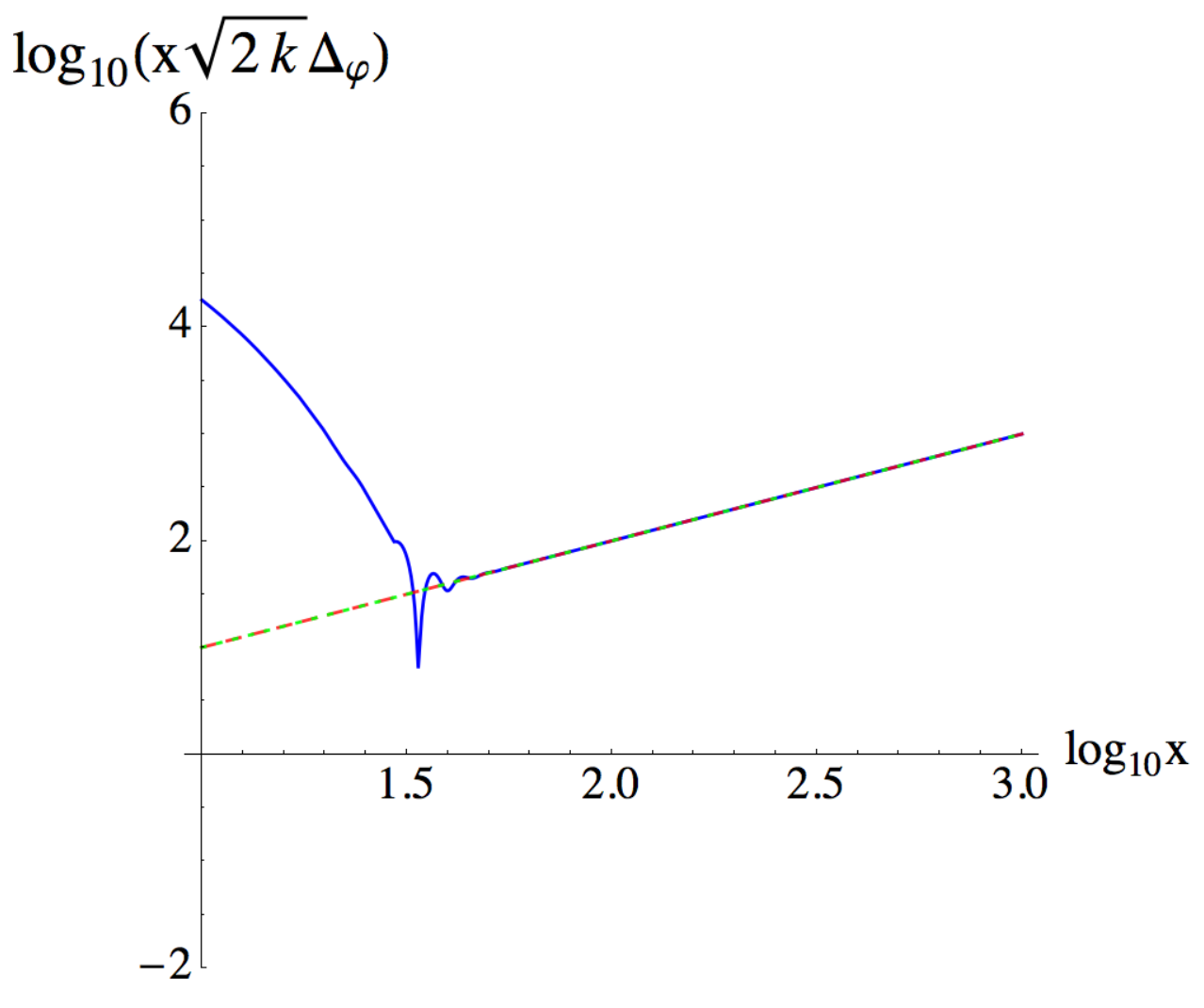}
\end{center}
\caption{In this plot, we show the time evolution of ~$x\sqrt{2k}~\Delta_\varphi$ ~with ~$m_Q =1$ ~(solid blue line), ~$m_Q =2$ ~(dashed red line) and ~$m_Q =3$ ~(dotted green line).
 The red and green lines are indistinguishable and stable, while the blue line experiences an exponential enhancement due to the instability.
 Note that we set ~$(\Lambda, \ |I_\varphi Q|) \approx ( 10^2, \ 10^{-4})$ ~in this plot.}
\label{fig: scalarfluc}
\end{figure}

 After elimination of $Y$, we can define the following canonical fields 
\begin{align}
\delta\varphi &= \dfrac{\Delta_\varphi}{a} \ , \quad \delta\sigma = \dfrac{\Delta_\sigma}{a} \ ,  \notag \\
\delta Q &= \dfrac{\Delta_Q}{\sqrt{2}aI} \ , \quad U = \dfrac{1}{\sqrt{2}kaI}\left( \dfrac{agQ}{k}\Delta_Q + \sqrt{\dfrac{k^2 + 2a^2g^2Q^2}{k^2}}\Delta_U \right) \ .
\end{align}
We derived the equations of motions for these scalar perturbations  in the Appendix \ref{scalar}.
 We are interested in the stability condition of scalar perturbations in the region ~$m_Q \gtrsim \mathcal{O}(1)$ ~where sizable chiral gravitational waves are produced (see the previous section).
 For the time window ~$m_Q \ll x \lesssim \Lambda$ , the equations of motions for perturbations at the leading order are given  by
\begin{align}
&\dfrac{d^2 \Delta_{\varphi}}{dx^2} + \left( 1 + \dfrac{4I_{\bar{\varphi}}^2Q^2}{x^2} \right)\Delta_\varphi \simeq 0 
                         \label{eq: s0}  \ , \\
&\dfrac{d^2 \Delta_{\sigma}}{dx^2} + \left(1 + \dfrac{\Lambda^2m_Q^2}{x^2}\right)\Delta_\sigma - \dfrac{5}{\sqrt{2}}\dfrac{\Lambda m_Q}{x}\dfrac{d \Delta_Q}{dx} - \dfrac{\sqrt{2}\Lambda}{x}\Delta_U \simeq 0 \label{eq: sS} \ , \\
&\dfrac{d^2 \Delta_{Q}}{dx^2} + \Delta_Q + \dfrac{5}{\sqrt{2}}\dfrac{\Lambda m_Q}{x}\dfrac{d \Delta_\sigma}{dx} \simeq 0 \label{eq: sQ} \ , \\
&\dfrac{d^2 \Delta_{U}}{dx^2} + \Delta_U - \dfrac{\sqrt{2}\Lambda}{x}\Delta_\sigma \simeq 0 \label{eq: sU} \ .
\end{align}
 Note that, for the coefficient of gauge-dilaton interactions ~$I_{\bar{\varphi}} Q$ , we assumed that the background parameter region satisfies ~$| I_{\bar{\varphi}} Q | \ll \Lambda$ , and neglected them in the above derivation.
 This assumption can be justified since ~$I$ ~needs to become small enough to get a large $\Lambda$ value, which is proportional to ~$I^{-1}$ .  
 Hence  dilaton perturbations do not affect the stability on sub-horizon scales.
 In order to check the stability condition of the scalar dynamics, we focus on one peculiar solution.
 We use the WKB method for analyzing these equations. Substituting 
\begin{equation}
\Delta_{\varphi} = C_1(x) e^{i S(x)} \ , \quad \Delta_{\sigma} = C_2(x) e^{i S(x)} \ , \quad \Delta_{Q} = C_3(x) e^{i S(x)} \ , \quad \Delta_{U} = C_4(x) e^{i S(x)} \ ,
\end{equation}
 into Eqs.(\ref{eq: s0}) - (\ref{eq: sU}),  we get
\begin{align}
& \left( 1 + \dfrac{4I_{\bar{\varphi}}^2Q^2}{x^2} - S_x^2 \right)C_1 \simeq 0 \ , \\
& \left(1 + \dfrac{\Lambda^2m_Q^2}{x^2} - S_x^2 \right)C_2 - \dfrac{5}{\sqrt{2}}\dfrac{\Lambda m_Q}{x}iS_x C_3 - \dfrac{\sqrt{2}\Lambda}{x}C_4 \simeq 0 \label{eq: sS2} \ , \\
&\left(1 - S_x^2 \right)C_3 + \dfrac{5}{\sqrt{2}}\dfrac{\Lambda m_Q}{x}i S_x C_2 \simeq 0 \label{eq: sQ2} \ , \\
&\left(1 - S_x^2 \right)C_4 - \dfrac{\sqrt{2}\Lambda}{x}C_2 \simeq 0 \label{eq: sU2} \ ,
\end{align}
where ~$S_x \equiv dS/dx = \omega$ ~is an angular frequency and we neglected the time dependence of ~$C_i(x)$.
 Let us focus on the solution with $C_1 = 0$.
 Then, from \eqref{eq: sS2} - \eqref{eq: sU2}, we get the relation
\begin{equation}
\left(1 + \dfrac{\Lambda^2m_Q^2}{x^2} - \omega^2 \right)\left(1 - \omega^2 \right) - \dfrac{25}{2}\dfrac{\Lambda^2 m_Q^2}{x^2}\omega^2 - \dfrac{2\Lambda^2}{x^2} = 0 \ .
\end{equation}
Thus, we have two modes
\begin{equation}
\omega_{\pm}^2 = \dfrac{1}{2x^2}\left(  2x^2 + \dfrac{29}{2}\Lambda^2m_Q^2\left( 1 \pm \sqrt{1 + \dfrac{8x^2}{29\Lambda^2m_Q^2}\left(\dfrac{25m_Q^2 + 2}{29m_Q^2}\right)} \right) \right) \label{eq: WKBsol1} \ .
\end{equation}
 Recalling that we are considering the region ~$x \lesssim \Lambda m_Q$, the plus mode ~$\omega_+^2$ ~is approximately given  by
\begin{equation}
\omega_+^2 \approx \dfrac{29\Lambda^2m_Q^2}{x^2} \ .
\end{equation}
 Apparently this mode is  positive and stable.
 On the other hand, for a minus mode $\omega_-^2$ by expanding the square root in \eqref{eq: WKBsol1} with respect to $x^2/\Lambda^2m_Q^2$ , we get
\begin{equation}
\omega_{-}^2 \approx \dfrac{2( 2m_Q^2 - 1 )}{29m_Q^2} \ .
\end{equation}
Thus, this mode has the instability in the region $m_Q < \sqrt{2}$ .
 We get the same result as in the previous works~\cite{Dimastrogiovanni:2012ew, Adshead:2013qp, Adshead:2013nka}, because the interaction of dilaton perturbations is not dominant in this stage.
 In order to confirm this instability, we numerically solved the full scalar equations of motions \eqref{eq: full1} - \eqref{eq: full4}.
 The time evolution of the scalar fluctuation ~$\Delta_\varphi$ ~with several ~$m_Q$ ~values are plotted in FIG.\ref{fig: scalarfluc}.
 We can see that in the region ~$m_Q < \sqrt{2}$ ~the instability actually occurs.
 Therefore we can avoid the instability by choosing ~$m_Q > \sqrt{2}$ ~in the late stage of inflation.
 Remarkably, in this parameter region, sizable chiral gravitational waves can be generated.

\begin{figure}[h]
\begin{center}
\includegraphics[width=8cm,height=8cm,keepaspectratio]{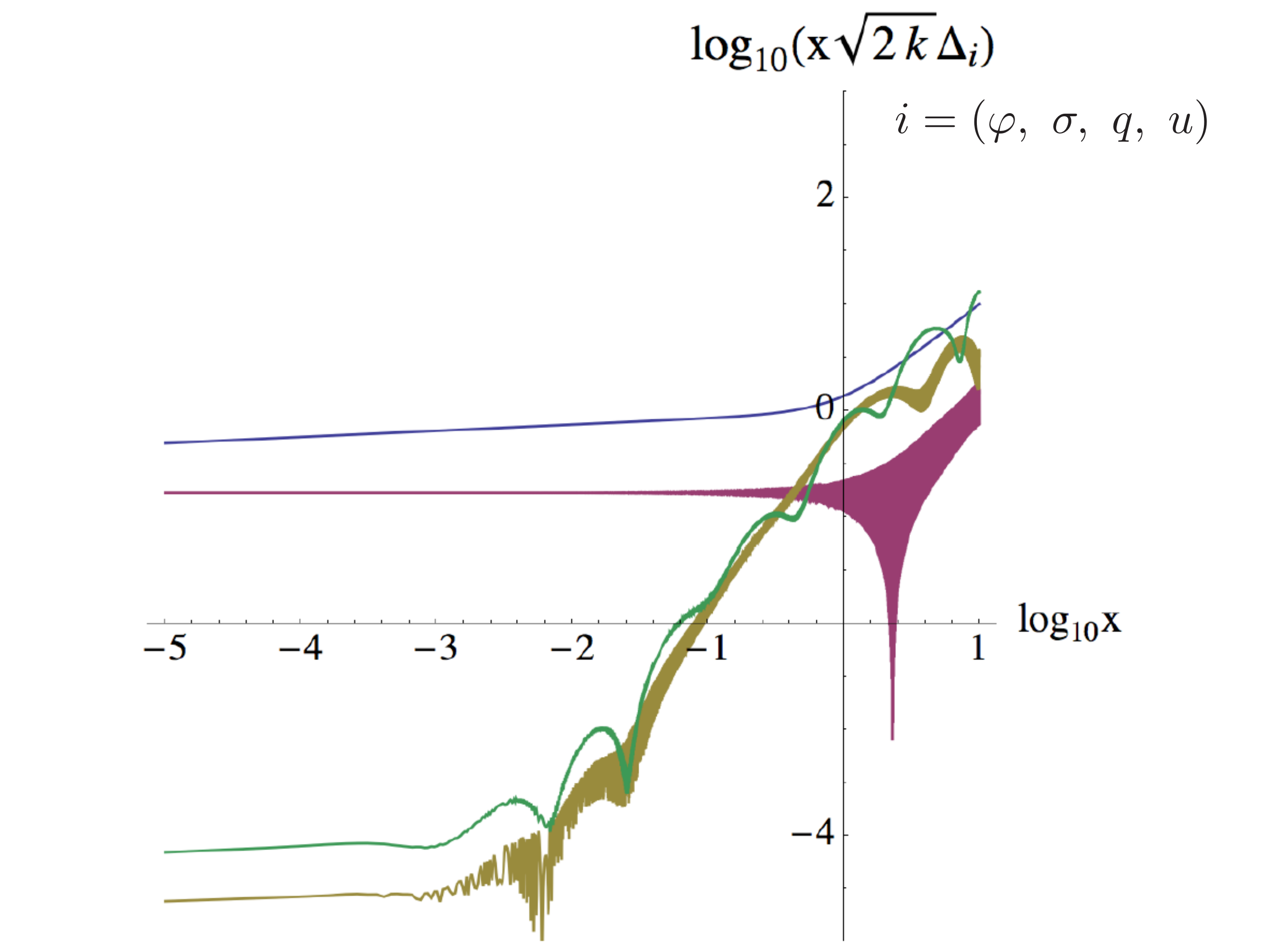}
\end{center}
\caption{We plotted the time evolution of ~$x\sqrt{2k}~\Delta_\varphi$ (blue line), ~$x\sqrt{2k}~\Delta_\sigma$ (red line), ~$x\sqrt{2k}~\Delta_q$ (yellow line) and ~$x\sqrt{2k}~\Delta_u$ (green line) ~in the region ~$x \lesssim m_Q$ .
 At the horizon-crossing, ~$x~\Delta_\varphi$ ~and ~$x~\Delta_\sigma$ ~stop decreasing and become constant in super-horizon due to the expansion of scale factor.
 At late times in super-horizon, this feeds back into the gauge field dynamics and makes their late time behavior constant.
 Note that we set ~$(m_Q, \ \Lambda, \ |I_\varphi Q|) \approx (2, \ 10^2, \ 10^{-4})$ in this plot.}
\label{fig: scalarsu}
\end{figure}

 The scalar dynamics in the region ~$x \lesssim m_Q$  have the similar feature as that of chromo-natural inflation \cite{Adshead:2013nka}.
 In this region, gauge fluctuations have no instability and ~$\Delta_\varphi$ ~and ~$\Delta_\sigma$ show the conventional power law behavior.
 We numerically solved the full scalar equations of motions \eqref{eq: full1} - \eqref{eq: full4} and plotted the scalar dynamics 
 in the region ~$x \lesssim m_Q$ ~in FIG.\ref{fig: scalarsu}.

\section{Discussion \label{phe}}

 In this section, we discuss constraints from the CMB observations and the possibility of generating chiral gravitational waves in this model. 
 
 In Sec.\ref{cur} , we estimated the curvature perturbation spectrum on CMB scales.
 Its dimensionless power spectrum is given by
\begin{equation}
\dfrac{k^3}{2\pi^2}\mathcal{P}_\zeta(k) \simeq \dfrac{H^2}{8\pi^2\epsilon_\varphi}\left( 1 + \mathcal{A}~\mathcal{E}_E^2~(\ln x_f)^2 \right) \label{eq:curs}
\end{equation}
at the end of inflation ~$x = x_f$ .
 Primordial fluctuations of CMB temperature are produced between ~$-60 \leq \ln x_f \leq -50$ , so the second term is dominant unless ~$\mathcal{E}_E^2$ ~is negligibly small. 
 The spectral index ~$n_s$ ~at the end of inflation is given by
\begin{equation}
n_s - 1 = \dfrac{d \ln{k^3\mathcal{P}_\zeta}}{d \ln k} \simeq \dfrac{2\mathcal{A}~\mathcal{E}_E^2~(\ln x_f)}{1 + \mathcal{A}~\mathcal{E}_E^2~(\ln x_f)^2} \label{eq: ns} \ .
\end{equation}
 This takes the value ~$2/(\ln x_f) \sim -0.04$ ~if the second term in the denominator in \eqref{eq: ns} is larger than the unity.
 Thus the spectral index will be nicely consistent with Planck data, even with ~$\mathcal{E}_E^2 \sim 1$ \cite{Ade:2015lrj}.
 Moreover, from the power spectrum of gravitational waves given in \eqref{eq: gwCMB} , we can estimate the tensor-to-scalar ratio as
\begin{equation}
r = \dfrac{\mathcal{P}^+_h + \mathcal{P}^-_h}{\mathcal{P}_\zeta} \simeq 16\epsilon_\varphi^2\dfrac{\epsilon_\varphi^{-1} + 4\mathcal{E}_E^2(\ln x_f)^2}{1 + \mathcal{A}~\mathcal{E}_E^2~(\ln x_f)^2} \label{eq: r}
\end{equation}
and the spectral tilt of tensor modes at the end of inflation
\begin{equation}
n_t = \dfrac{d \ln{k^3\mathcal{P}^\pm_h}}{d \ln k} \simeq \dfrac{8\epsilon_E~(\ln x_f)}{1 + 4\epsilon_E~(\ln x_f)^2} \label{eq: nt} \ .
\end{equation}
 Note that we neglected the contribution of ~$\epsilon_B$ ~in \eqref{eq: r} and \eqref{eq: nt} since its coefficient ~$m_Q$ ~is sufficiently small in this epoch.
 Remarkably, ~$r$ ~can be suppressed by a factor ~$\epsilon_\varphi/\mathcal{A}$ ~compared with the conventional single field slow-roll inflation.
 Moreover, ~$n_t$ ~is also red but much closer to the scale invariance than that of scalar perturbations if ~$\mathcal{E}_E^2$ ~is sufficiently small.
 Therefore, without any difficulty, we can choose appropriate ~$\mathcal{E}_E^2$ 
and make the model consistent with the  CMB constraints.
 Though we have not calculated the bispectrum of curvature perturbations and check the qualitative features of non-gaussianity, we expect that its constraint is not so strong unless the axion-gauge interactions are dominant on CMB scales \cite{Funakoshi:2012ym}.

 The most remarkable prediction in this model is that primordial chiral gravitational waves are generated on small scales, 
which happens after $25$ e-folds in our numerical example (see FIG.\ref{fig: gauge}).
 The frequencies of gravitational waves corresponding to CMB scales are around $10^{-18}~\text{Hz}$ ~today \cite{Maggiore:1999vm}, so chiral gravitational waves in our model can be detectable by gravitational wave interferometers whose sensitivity peaks are located above $\text{nHz}$ ~frequency regions .
 From the discussion in Sec.\ref{pe}, the ratio of the amplitude of gravitational wave spectrum is approximately given by 
\begin{eqnarray}
\dfrac{\mathcal{P}^+_h(k)}{\mathcal{P}^-_h(k)} \simeq
\begin{cases}
 1  \qquad (f \lesssim \text{nHz}) & \\
 1 + 8 |C_2|^2Q_{\text{min}}^2\left| \mathcal{I}_0(m_{Q}) \right. - m_{Q}\mathcal{I}_1(m_{Q}) \left.+ m_{Q}^2 \mathcal{I}_2(m_{Q}) \right|^2 \qquad (f \gtrsim \text{nHz}) \ , &
\end{cases}
\end{eqnarray}
where ~$f = \tfrac{k}{2\pi}$ ~is the frequency of gravitational waves today.
 We expect that this two branches are smoothly connected at the transition epoch of two inflationary stages ($f \sim \text{nHz}$).

 Finally, let us estimate the intensity of gravitational waves in this model.
 The intensity derived for vacuum fluctuations is given by \cite{Maggiore:1999vm}
\begin{equation}
h_0^2\Omega_{\text{vac}}(f) \simeq 10^{-13}\left(\dfrac{H}{10^{-4}}\right)^2 \ .
\end{equation}
 Considering CMB normalization in \eqref{eq:curs}, Hubble parameter can be estimated as ~$10^{-7} \lesssim H \lesssim 10^{-5}$ for certain ~$\mathcal{E}_E^2$ ~region, ~then ~$h_0^2\Omega_{\text{vac}}$ ~satisfies $10^{-19} \lesssim h_0^2\Omega_{\text{vac}}(f) \lesssim 10^{-15}$ .
 In the late stage of inflation, the enhancement of chirality from the vacuum case is like FIG.\ref{fig: CGWratio}, getting large value as ~$m_Q$  increases, so there are some sets of parameter region ~$(Q_{\text{min}}, \ m_Q)$ ~where chiral gravitational waves can be observed by various detectors such as DECIGO ($h_0^2\Omega_{\text{GW}} \gtrsim 10^{-20}$) \cite{Seto:2001qf}, eLISA ($h_0^2\Omega_{\text{GW}} \gtrsim 10^{-10}$) \cite{AmaroSeoane:2012km} and SKA ($h_0^2\Omega_{\text{GW}} \gtrsim 10^{-14}$) \cite{Carilli:2004nx}.

\section{Conclusion \label{con}}

 In this paper, we studied an inflationary dynamics with a dilaton and an axion coupled to a SU(2) gauge field.
 Specifically, inspired by supergravity, we introduced a non-trivial gauge kinetic function to  chromo-natural inflation~\cite{Adshead:2012kp}.
 Consequently,  the effective coupling constant of the axion to the gauge field is dynamically controlled.
 For the initial conditions, we required that the gauge field is near the origin of its effective potential and the dilaton 
 playing a role of inflaton initially is energetically dominant.
 Since the interaction of the axion to the gauge field is sufficiently small, 
 we could neglect the axion in the background dynamics on CMB scales.
 We derived the power-spectrum of scalar and tensor perturbations in this early stage of inflation by using the in-in formalism, 
 and discussed its observational implication by looking at the spectral index and the tensor-to-scalar ratio.
 We found that they are characterized by the kinetic energy density of the gauge field and
 can be controlled so that the model becomes consistent with the constraints from the current CMB observations.
 In particular, since the-gauge field is inert in the early stage of inflation,
  the model does not show any conflict with the CMB constraints.

 The main prediction of this paper is the generation of chiral gravitational waves on small scales.
 We found an inflationary solution where the gauge field grows due to the gauge-kinetic function and finally settles in a finite value ~$Q_{\text{min}}$ , which realizes the delayed chromo-natural inflation.
 Then we analyzed tensor dynamics in the late stage of inflation and computed gravitational wave spectrum by employing the in-in formalism, and got 
 the qualitatively similar result as chromo-natural inflation \cite{Adshead:2013nka}.
 The general feature of tensor perturbations is the enhancement of chiral gravitational waves which 
 is controlled by an attractor value ~$Q_{\text{min}}$ ~and the mass parameter of the gauge field ~$m_Q$ .
 Moreover, we checked the stability condition of scalar dynamics and showed that there is a stable parameter region ~$m_Q > \sqrt{2}$ .
 Note that during this period we have chosen the  gauge-kinetic function so that the interaction of the dilaton to the gauge field does not contribute to perturbation dynamics.
 We found that there is a parameter region where chiral gravitational waves are produced in an interesting frequency range, higher than nHz, which might be detectable in future gravitational wave interferometers and pulsar timing arrays, such as DECIGO, eLISA and SKA.
 
 It is interesting to investigate the reheating stage in this model.
 We expect from our previous work \cite{Obata:2014loa} that at this epoch the effective mass parameter ~$m_Q$ ~grows and then chiral 
 gravitational waves are more enhanced, which might reach the current observational bounds on ground-based interferometers.
 In this paper, we studied  our model with an isotropic initial condition.
Notice that the anisotropy remains in models with a gauge kinetic function for a non-Abelian gauge field \cite{Murata:2011wv}.
 While, in the case of chromo-natural inflation, the initial anisotropy decays and the isotropic configuration is stable \cite{Maleknejad:2011jr}.
 Thus, it is intriguing to investigate the fate of the  initial anisotropy in our model.
 We leave these issues for future work.

\section*{Acknowledgements}

 We would like to thank P. Adshead and M. Wyman for fruitful notebook and discussion.
This work was supported by  JSPS KAKENHI Grant Number 25400251,
 MEXT  KAKENHI Grant Number 26104708, and MEXT KAKENHI Grant Number 15H05895. 

\appendix

\section{Equations of motion for scalar perturbations \label{scalar}}

In this appendix, we list up the full equations of motion used for numerical calculations.
 We set ~$\phi = \beta = 0$ .
 Then, we have the following equations of motions for scalar perturbations 
\begin{align}
&\dfrac{d^2 \Delta_{Q}}{dx^2} + \mathcal{Q}_1\Delta_Q + \mathcal{Q}_2\Delta_U + \mathcal{Q}_3\dfrac{d \Delta_\varphi}{dx} + \mathcal{Q}_4\Delta_\varphi + \mathcal{Q}_5\dfrac{d \Delta_\sigma}{dx} + \mathcal{Q}_6\Delta_\sigma = 0 \label{eq: full1} \ , \\
&\dfrac{d^2 \Delta_{U}}{dx^2} + \mathcal{U}_1\Delta_U + \mathcal{U}_2\Delta_Q + \mathcal{U}_3\dfrac{d \Delta_\varphi}{dx} + \mathcal{U}_4\Delta_\varphi + \mathcal{U}_5\dfrac{d \Delta_\sigma}{dx} + \mathcal{U}_6\Delta_\sigma = 0 \label{eq: full2} \ , \\
&\dfrac{d^2 \Delta_{\varphi}}{dx^2} + \mathcal{P}_1\Delta_\varphi + \mathcal{P}_2\Delta_\sigma + \mathcal{P}_3\dfrac{d \Delta_Q}{dx} + \mathcal{P}_4\Delta_Q + \mathcal{P}_5\dfrac{d \Delta_U}{dx} + \mathcal{P}_6\Delta_U = 0 \label{eq: full3} \ , \\
&\dfrac{d^2 \Delta_{\sigma}}{dx^2} + \mathcal{S}_1\Delta_\sigma + \mathcal{S}_2\Delta_\varphi + \mathcal{S}_3\dfrac{d \Delta_Q}{dx} + \mathcal{S}_4\Delta_Q + \mathcal{S}_5\dfrac{d \Delta_U}{dx} + \mathcal{S}_6\Delta_U = 0 \label{eq: full4} \ .
\end{align}
 As for the coefficients ~$\mathcal{Q}_i \ (i=1, ... , 6)$ , we get
\begin{align}
\mathcal{Q}_1&= \dfrac{1}{x^4(x^2+2m_Q^2)}\left(4m_Q^6 + 2m_Q^4 \left(2 - 2d\ln{I}/dx ~x + 5x^2 \right) + \left( 1 - (d\ln{I}/dx)^2 \right)x^6 \right. \notag \\
 &\left. + 2m_Q^2 x^2\left( 1 - d\ln{I}/dx ~x + \left( 3 - (d\ln{I}/dx)^2 \right)x^2 \right) \right. \notag \\
 &\left. - 2\xi\left(2m_Q^5 + 5m_Q^3 x^2 + 2m_Q x^4\right) \right) - \dfrac{d^2 I/dx^2}{I} + \dfrac{(d I/dx)^2}{I^2} \ , \\
\mathcal{Q}_2&= \dfrac{1}{x^4(x^2 + 2m_Q^2)^{3/2}}\left( 8m_Q^7 + 8m_Q^5( 1 - d\ln{I}/dx ~x ) +2m_Q^3(4 - 4d\ln{I}/dx ~x - 3x^2)x^2  \right. \notag \\
 &\left. + 2m_Q(1 - d\ln{I}/dx ~x - x^2 )x^4 - \xi(8m_Q^6 + 4m_Q^4x^2 - 2m_Q^2x^4 - x^6 ) \right) \ , \\
\mathcal{Q}_3&= -\dfrac{2\sqrt{2}I_{\bar{\varphi}} Q}{x} \ , \\
\mathcal{Q}_4&= \dfrac{2\sqrt{2}I_{\bar{\varphi}}Q}{x^2}\left( 1 - d\ln{I}/dx ~x + 2m_Q^2 \right) - \dfrac{2\sqrt{2}Q d I_{\bar{\varphi}}/dx}{x} \ , \\
\mathcal{Q}_5&= \dfrac{5}{\sqrt{2}}\dfrac{\lambda}{I f}\dfrac{Qm_Q}{x} \ , \\
\mathcal{Q}_6&= \dfrac{5}{\sqrt{2}}\dfrac{\lambda}{I f}\dfrac{Qm_Q}{x^2} \ .
\end{align}
 As for ~$\mathcal{U}_i \ (i=1, ... , 6)$, we have
\begin{align}
\mathcal{U}_1&= \dfrac{1}{x^4(x^2+2m_Q^2)^2}\left(16m_Q^8 + 16m_Q^6 \left(1 - d\ln{I}/dx ~x + 2x^2 \right) + \left( 1 - (d\ln{I}/dx)^2 \right)x^8 \right. \notag \\
 &\left. + 4m_Q^4 x^2\left( 6(1 - d\ln{I}/dx ~x) + \left( 6 - (d\ln{I}/dx)^2 \right)x^2 \right) \right. \notag \\
 &\left. + 2m_Q^2 x^4\left( 7 - 4d\ln{I}/dx ~x + 2\left( 2 - (d\ln{I}/dx)^2 \right)x^2 \right) \right. \notag \\
 &\left. - 2\xi(8m_Q^7+12m_Q^5 x^2+6m_Q^3 x^4 + m_Q x^6) \right) - \dfrac{d^2 I/dx^2}{I} + \dfrac{(d I/dx)^2}{I^2} \ , \\
\mathcal{U}_2&= \mathcal{Q}_2 \ , \\
\mathcal{U}_3&= \dfrac{2\sqrt{2}I_{\bar{\varphi}}Q m_Q}{x(x^2 + 2m_Q^2)^{1/2}} \ , \\
\mathcal{U}_4&= -\dfrac{2\sqrt{2}I_{\bar{\varphi}}Q m_Q}{x^2(x^2 + 2m_Q^2)^{3/2}}\left( 4m_Q^4 + 2m_Q^2(1 - d\ln{I}/dx ~x - d\ln{I_{\bar{\varphi}}}/dx ~x + 2x^2) \right. \notag \\
 &\left. + (3 - d\ln{I}/dx ~x - d\ln{I_{\bar{\varphi}}}/dx ~x + x^2)x^2 \right) \ , \\
\mathcal{U}_5&= -\dfrac{\lambda}{I f}\dfrac{\sqrt{2}Qm_Q^2}{x(x^2 + 2m_Q^2)^{1/2}} \ , \\
\mathcal{U}_6&= -\dfrac{\lambda}{I f}\dfrac{\sqrt{2}Q}{x^2(x^2+m_Q^2)^{3/2}}\left( 2m_Q^4 + m_Q^2 x^2 + x^4 \right) \ .
\end{align}
 Then, as for ~$\mathcal{P}_i \ (i=1, ... , 6)$, we have
\begin{align}
\mathcal{P}_1&= 1 + \dfrac{1}{x^2}\left( 3\left(\dfrac{I_{\bar{\varphi}}^2}{I^2} + \dfrac{I_{\bar{\varphi}\bar{\varphi}}}{I} \right)(\epsilon_B - \epsilon_E) - 2 + \dfrac{V_{\bar{\varphi}\bar{\varphi}}}{H^2} + \dfrac{4I_{\bar{\varphi}}^2Q^2}{x^2 + 2m_Q^2}x^2 \right) \ , \\
\mathcal{P}_2&= -\dfrac{\lambda}{I f}\dfrac{2I_{\bar{\varphi}} Q^2m_Q}{x^2 + 2m_Q^2} \ , \\
\mathcal{P}_3&= -\mathcal{Q}_3 \ , \\
\mathcal{P}_4&= \dfrac{2\sqrt{2}I_{\bar{\varphi}}Q}{x^2}\left( 2m_Q^2 - d\ln{I}/dx ~x \right) \ , \\
\mathcal{P}_5&= -\mathcal{U}_3 \ , \\
\mathcal{P}_6&=  -\dfrac{2\sqrt{2}I_{\bar{\varphi}}Q m_Q}{x^2(x^2 + 2m_Q^2)^{3/2}}\left( 4m_Q^4 + 2m_Q^2( - d\ln{I}/dx ~x + 2x^2) + (1 - d\ln{I}/dx ~x + x^2)x^2 \right) \ .
\end{align}
 Finally, as for ~$\mathcal{S}_i \ (i=1, ... , 6)$, we have
\begin{align}
\mathcal{S}_1&= 1 + \dfrac{1}{x^2}\left( -2 + \dfrac{W_{\bar{\sigma}\bar{\sigma}}}{H^2} + \dfrac{\alpha^2}{I^2 f^2}\dfrac{Q^2m_Q^2}{x^2 + 2m_Q^2}x^2 \right) \ , \\
\mathcal{S}_2&= \mathcal{P}_2 \ , \\
\mathcal{S}_3&= -\mathcal{Q}_5 \ , \\
\mathcal{S}_4&= \dfrac{10}{\sqrt{2}}\dfrac{\alpha}{I f}\dfrac{Qm_Q}{x^2}\left( 1 + \dfrac{1}{2}d\ln{I}/dx ~x \right) \ , \\
\mathcal{S}_5&= -\mathcal{U}_5 \ , \\
\mathcal{S}_6&=  -\dfrac{\lambda}{I f}\dfrac{\sqrt{2}Q}{x^2(x^2+2m_Q^2)^{3/2}}\left( 2m_Q^4(2 + d\ln{I}/dx ~x) + m_Q^2(3 + d\ln{I}/dx ~x)x^2 + x^4 \right) \ .
\end{align}

\end{document}